\documentclass[a4paper,11pt,headings=big,DIV=12]{scrartcl}
\pdfoutput=1

\usepackage{array} 
\usepackage{booktabs} 
\usepackage{caption}

\usepackage[a4paper, top=1.2in, bottom=1.2in, left=1.1in, right=1.1in]{geometry}
\usepackage[utf8]{inputenc}
\usepackage{amsmath, amssymb}
\usepackage{color}
\usepackage{comment}
\usepackage[dvipsnames, table]{xcolor}
\definecolor{blue}{rgb}{0,0,0.5}
\usepackage[colorlinks,linkcolor=blue,urlcolor=blue,citecolor=blue]{hyperref}
\usepackage{graphicx}
\graphicspath{{./figs/}}
\usepackage{slashed}
\usepackage{cite}
\usepackage{enumitem}
\usepackage{bm}
\usepackage{physics}
\usepackage{xspace}
\usepackage[T1]{fontenc}
\usepackage{lmodern}

\newcommand{\be}{\begin{equation}}
\newcommand{\ee}{\end{equation}}
\newcommand{\bea}{\begin{eqnarray}}
\newcommand{\eea}{\end{eqnarray}}

\newcommand{\mc}{\mathcal}

\newcommand{\nn}{\nonumber}
\newcommand{\noi}{\noindent}

\def\max{{\rm max}}
\def\min{{\rm min}}

\newcommand{\TeV}{{\rm TeV}}
\def\refeq#1{Eq.~(\ref{#1})}

\def\reftab#1{Table~\ref{#1}}
\def\reffig#1{Fig.~\ref{#1}}
\def\reffigs#1#2{Figs.~\ref{#1}-\ref{#2}}
\def\refsec#1{Sec.~\ref{#1}}

\def\de{\partial}
\def\diag{{\rm diag}}

\def\hc{{\rm h.c.}}
\def\re{{\rm Re}}
\def\im{{\rm Im}}

\newcommand{\MK}{m_K}
\newcommand{\MPI}{m_\pi}
\newcommand{\al}{\alpha}
\newcommand{\De}{\Delta}
\newcommand{\DeK}{\Delta_K}
\newcommand{\DePI}{\Delta_\pi}
\newcommand{\Ks}{K_S}
\newcommand{\BR}{\mc{B}}
\newcommand{\Lst}{{\mc L_{{\rm st}}}}
\newcommand{\mcM}{{\mc M}}

\newcommand{\cgg}{c_{GG}}

\DeclareOldFontCommand{\rm}{\normalfont\rmfamily}{\mathrm}
\DeclareOldFontCommand{\sf}{\normalfont\sffamily}{\mathsf}
\DeclareOldFontCommand{\tt}{\normalfont\ttfamily}{\mathtt}
\DeclareOldFontCommand{\bf}{\normalfont\bfseries}{\mathbf}
\DeclareOldFontCommand{\it}{\normalfont\itshape}{\mathit}
\DeclareOldFontCommand{\sl}{\normalfont\slshape}{\@nomath\sl}
\DeclareOldFontCommand{\sc}{\normalfont\scshape}{\@nomath\sc}

\newcommand{\ecker}{Ecker:1987qi}
\newcommand{\scherer}{Scherer:2002tk}
\newcommand{\georgi}{Georgi:1986df}
\newcommand{\neubertPRL}{Bauer:2021wjo}
\newcommand{\neubertALPslong}{Bauer:2021mvw}
\newcommand{\alves}{AlvesJunior:2018ldo}
\newcommand{\pdg}{ParticleDataGroup:2024cfk}

\newcommand{\adler}{E787:2000iwe}
\newcommand{\epimu}{\varepsilon_{\pi \mu}}
\newcommand{\emupi}{\varepsilon_{\mu \pi}}
\newcommand{\pipi}{\pi \pi}
\newcommand{\mumu}{\mu \mu}
\newcommand{\alem}{\alpha_{\rm em}}
\newcommand{\BRth}{\BR_{\rm th}}
\newcommand{\Seff}{S_{\rm eff}}
\newcommand{\pippim}{\pi^+ \pi^-}
\newcommand{\mupmum}{\mu^+ \mu^-}
\newcommand{\AKL}{A_{K_L}}

\begin{document}

\begin{flushright}
\small
LAPTH-059/24
\end{flushright}
\vskip0.5cm

\begin{center}
{\sffamily \bfseries \LARGE \boldmath
Probing QCD Axions or Axion-like Particles\\[0.3cm]in three-body $K$ Decays}
\\[0.8 cm]
{\normalsize \sffamily \bfseries Maël Cavan-Piton$^1$, Diego Guadagnoli$^1$, Axel Iohner$^1$, Diego Martínez Santos$^2$, Ludovico Vittorio$^1$} \\[0.5 cm]
\small
$^1${\em LAPTh, Universit\'{e} Savoie Mont-Blanc et CNRS, Annecy, France}\\
[0.1cm]
$^2${\em 
Ferrol Industrial Campus, Dr. Vázquez Cabrera, s/n, 15403, Universidade de A Coruña, Spain
}\\
\end{center}

\medskip

\begin{abstract}
\noi Two-body decays like $K \to \pi a$ rank among the most constraining collider probes for new, low-mass, feebly interacting pseudoscalar particles $a$. We explore an alternative class of kaon decay modes, specifically three-body decays to $\pi \pi a$ or $\mu \mu a$. The former occur at tree level, while the latter is loop-suppressed yet accidentally finite. These modes specifically leverage the accurate tracking capabilities at LHCb. We present an estimation approach for the sensitivity achievable in future searches within these channels. Our argument uses the current {\em uncertainty} in leading irreducible backgrounds identified for each channel.
Our findings suggest that dedicated searches could probe $f_a$ scales between $10^4$ and $10^6$ TeV, highlighting their strong potential. A direct comparison with actual searches, only available in the $K^+ \to \pi^+ \pi^0 a$ channel, supports this conclusion. Finally, we show that, in these searches, reconstruction efficiency maps are such that large efficiencies are naturally aligned with regions of higher yields in Dalitz plots.
\end{abstract}

\section{Introduction}

New low-mass pseudoscalar particles are among the most sought-after realizations of physics beyond the Standard Model, for several reasons. First, such particles, even with masses in the MeV-GeV range and larger-than-weak coupling strengths to ordinary matter, may help address fundamental problems, ranging from dark matter and the strong $CP$ problem to mass hierarchies~\cite{Lanfranchi:2020crw}, while remaining entirely compatible with our extensive knowledge of stable matter. In fact, the strongest of these constraints mostly come from astrophysical data, and thus apply primarily to interactions with 1st-generation matter~\cite{Kolb:1990vq}---although it has recently been shown that sufficiently hot transients, such as supernov\ae{}, constitute an exception~\cite{Cavan-Piton:2024ayu}, as they also constrain interactions with the strange quark. Second, there is no first-principle reason why such new particles should couple with the same strength with all generations of matter, or in a flavour-diagonal way~\cite{\georgi}. In fact, their couplings are not constrained by known symmetries such as gauge invariance. Hence these particles, if they exist, are expected to couple to the heavier generations of matter as well. Meson (or tau-lepton) decays at colliders are well-suited to test such couplings, considering the large statistics and accuracy now achievable at flavour facilities.

The traditionally strongest probes of flavour-violating axion couplings include two-body decays such as $K \to \pi a$. In this study, we explore a different set of modes, namely three-body decays of the kind $K_S \to h^+ h^- a$, where $h = \pi$ or $\mu$. Other aspects of three-body $K$ decays have been recently discussed in Refs.~\cite{Goudzovski:2022vbt,Alves:2020xhf,DiLuzio:2023cuk}. The main advantages of such modes are threefold: 
\begin{itemize}

\item The pionic channel $K_S \to \pi^+ \pi^- a$ occurs at tree level already, and allows to access the {\em axial} flavour-violating axion coupling to light quarks, namely $(k_A)_{23}$;

\item The muonic channel $K_S \to \mu^+ \mu^- a$, although loop-suppressed (as is the case for $K \to \mu^+ \mu^- \pi$ \cite{\ecker}), turns out to be accidentally {\em finite} (i.e. divergences cancel). This entails a very accurate theory prediction---and one that probes axion interactions in a chiral-loop process;

\item In the case of neutral kaons, we focus on the $K_S$, as the initial state needs to decay often enough within the detector volume. From  an experimental point of view, $h^+ h^-$ represent a very clean trigger, that exploits specific strengths at facilities like LHCb, such as the accurate tracking capabilities.\footnote{%
This point also leverages the very high statistics that LHCb can achieve on kaons produced and sufficiently short-lived to decay often enough within the detector volume.
LHCb has collected $9.56~{\rm fb}^{-1}$ of integrated luminosity in 2024, which should allow for sensitivities at the $10^{-9}$ level for $K_S \to \mupmum X$ branching fractions.
}

\end{itemize}

The theoretical predictions for these decay modes should be compared to dedicated experimental searches, most of which are yet to be performed. This makes a quantitative estimate of the achievable bounds not straightforward. However, one may estimate the sensitivity to be expected on such searches from existing data. This sensitivity can then be interpreted as an estimate of the bound one may realistically obtain in a given dedicated search. The basic idea is to identify the dominant backgrounds in a given search and then use the quoted {\em uncertainty} on such measurements as a basis for evaluating the sensitivity that the search can achieve. For instance, in an LHCb search for $K_S \to \pi^+ \pi^- a$, a dominant background would come from $K_S \to \pi^+ \pi^- \gamma$ for a light axion ($m_a \ll m_\pi$), and from $K_L \to \pi^+ \pi^- \pi^0$ for an axion with mass comparable to the pion's---despite the small $K_L$ acceptance at LHCb. The current uncertainty on these measurements is our starting point to estimate the sensitivity achievable in a dedicated search of $K_S \to \pi^+ \pi^- a$. Intuitively, the error on the dominant backgrounds quantifies the leeway for the signal decay. The detailed argument will be presented in \refsec{sec:num_res}.
We will also include the $K^+ \to \pi^+ \pi^0 a$ modes. Although these are outside the scope of LHCb, they offer remarkable sensitivity, that we will assess by performing an explicit comparison with a recent search.

We will use the sensitivity estimates for the different modes ($K_S \to \pi \pi a$,  $K_S \to \mu \mu a$, $K^+ \to \pi \pi a$) to assess the constraining power of a dedicated search on the fundamental axion-quark couplings mediating each decay. To do this, we will employ relations that equate the theoretically calculated total branching ratio (BR) with the sensitivity estimated as outlined above. In the plane of the real vs. imaginary parts of the couplings, these relations will help determine the maximum testable value of $f_a$ for a given decay, assuming the fundamental coupling is as large as theoretically permitted.
We will also discuss singly and doubly differential branching ratios (BRs). As concerns the latter, we will note that most of the considered $K_S$ modes are in practice dominated by small portions of the Dalitz plot. A well-defined way to improve the accuracy of the analysis is thus to optimize the efficiency maps to the very same regions of the Dalitz plot.

The plan of this work is as follows. \refsec{sec:TH} presents our theory calculations. In particular, in \refsec{sec:TH_setup} we summarize the theory framework in which our calculations are performed; applications to $K \to \pi \pi a$ and $K \to \ell \ell a$ (and contact with the reference $K \to \ell \ell \pi$ case) are discussed in \refsec{sec:Kpipia} and \ref{sec:Klla}, respectively. These theory predictions are subsequently compared with suitable experimental data in \refsec{sec:num_res}, to infer the sensitivity on the axial axion-down-strange coupling achievable in the different channels. We conclude in \refsec{sec:outlook}.

\section{\boldmath Theory aspects} \label{sec:TH}

\subsection{\boldmath Setup and Conventions} \label{sec:TH_setup}

\noi As elucidated in Ref.~\cite{\georgi}, axion-hadron interactions can be consistently inferred as a generalization of low-energy QCD, or Chiral Perturbation Theory (ChPT), to the extent that the axion is the near-Goldstone boson of a global $U(1)$ symmetry, precisely as pions are near-Goldstones of the non-singlet axial part of the QCD chiral symmetry. The effect of strangeness ($S$) changing weak interactions is included as a perturbation to the strong Lagrangian, and the axion is taken into account following the same logic as in Ref.~\cite{\georgi}, see Refs.~\cite{\neubertPRL,\neubertALPslong}.

Similarly as Ref. \cite{\ecker}, the interaction terms we necessitate fall into two categories: the ``strong'' Lagrangian, including mesonic couplings to external e.m. currents (here collectively indicated as $\mc L_{\rm st}$); the $|\Delta S| = 1$ terms enhanced by the $\Delta I = 1/2$ rule, including the induced e.m. penguins (here collectively indicated as $\mc L_{|\Delta S| = 1}$). Both sets of terms, $\mc L_{\rm st} + \mc L_{|\Delta S| = 1}$, need to incorporate the axion~\cite{\neubertPRL,\neubertALPslong}.\footnote{We accordingly follow the conventions in Refs.~\cite{\neubertPRL,\neubertALPslong}, unless they are different (e.g. in the chiral transformation properties of the meson field) than in Scherer's review \cite{\scherer}, in which case we comply with the latter.}
In this notation, these terms read
\begin{align}
\label{Lstrong}
&\mc L_{\rm st} = \frac{F_0^2}{4} \Bigl( \Tr D_\mu U (D^\mu U)^\dagger + 2 B_0 \Tr(M_q U^\dagger + U M_q^\dagger ) \Bigl)~,\\
\label{LDeltaS1}
&\mc L_{|\Delta S| = 1} = + \frac{4 G_F}{\sqrt2} V_{us} V_{ud}^* \, g_8 \left( L_\mu L^\mu \right)^{32} + \hc
\end{align}
with $M_q = \diag(m_u, m_d, m_s)$, $U \equiv \exp(i \frac{\phi^a}{F_0} \lambda^a)$ normalized such that $(\phi^a \lambda^a)_{11} = \pi^0 + \eta_8 / \sqrt3$, and transforming as $U \rightarrow R U L^\dagger$. The covariant derivative is given by
\be
\label{eq:DmuU}
D_\mu U = \partial_\mu U - i \frac{\partial_\mu a}{f_a} \left( \hat k_R(a) \, U - U \, \hat k_L(a) \right) - i e A_\mu \left[ Q, U \right]~,
\ee
where
\be
\label{eq:hatkRL}
\hat k_{R,L}(a) = e^{i \varphi^{\pm} a / f_a} (k_{R,L} + \varphi^\pm) e^{- i \varphi^{\pm} a / f_a}
\ee
with $Q = \diag(Q_u, Q_d, Q_s)$, $\varphi^\pm = c_{GG} (\delta \pm \kappa)$ and $\cgg$ denoting the axion-gluon coupling in the normalization $\mc L_{a} \supset \cgg \frac{\alpha_s}{4 \pi} \frac{a}{f_a} G_{\mu \nu} \tilde G^{\mu \nu}$, with $\tilde G_a^{\mu \nu} \equiv \frac{1}{2}\epsilon^{\mu \nu \rho \sigma} G_{\rho \sigma, a}$~\cite{\neubertPRL}. The matrices $\delta, \kappa$ are defined as $\delta = \diag(\delta_1, \delta_2, \delta_3)$, $\delta_i \in {\rm I\!R}$ and $\kappa = M_q^{-1} / \Tr(M_q^{-1})$, where $M_q$ is the light-quark mass matrix \cite{\georgi,\neubertPRL}. The dependence on contributions proportional to the $\delta$ and $\kappa$ parameters must cancel in physical observables \cite{\neubertPRL}.
The $\hat k_{R,L}$ couplings in \refeq{eq:hatkRL} encode the strength of the fundamental interactions between the axion and light quarks, with Lagrangian
\be
\label{eq:Laqq}
\mc L_{aqq} ~\equiv~ \frac{\de_\mu a}{f_a} 
\left(
\bar q \, \gamma_L^\mu \hat k_{L}{(a)} \, q + 
\bar q \, \gamma_R^\mu \hat k_{R}{(a)} \, q
\right)~,
\ee
with $q = (u, d, s)^T$ and $\gamma_{L,R}^\mu = \gamma^\mu(1\mp\gamma_5)/2$. As we consider processes involving a single axion, we can replace $\hat k_{L,R}(a) \to k_{L,R}$. Finally, the $L_\mu$ current appearing in \refeq{LDeltaS1} reads
\be
L_\mu^{ji} = i \frac{F_0^2}{2} e^{i a / f_a (\varphi^-_i - \varphi^-_j)} \left[ (D_\mu U)^\dagger U \right]^{ji}~,
\ee
where $\varphi^{\pm}_k = (\varphi^{\pm})_{kk}$.

\subsection{\boldmath Application to $K \to \pi^+ \pi^- a$}\label{sec:Kpipia}

The perhaps simplest example of our decays of interest is $K \to \pi^+ \pi^- a$.
\begin{figure}[h!]
 \centering
 \includegraphics[width=0.30\textwidth]{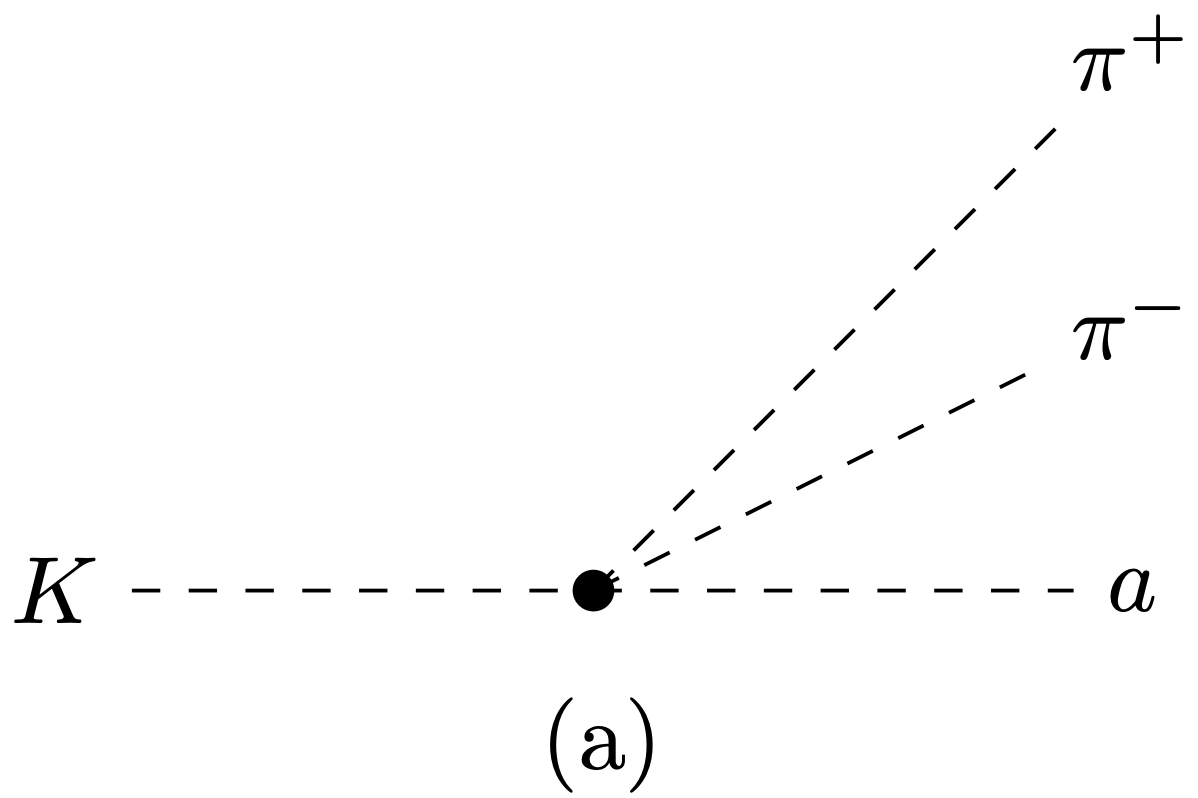}
 \includegraphics[width=0.30\textwidth]{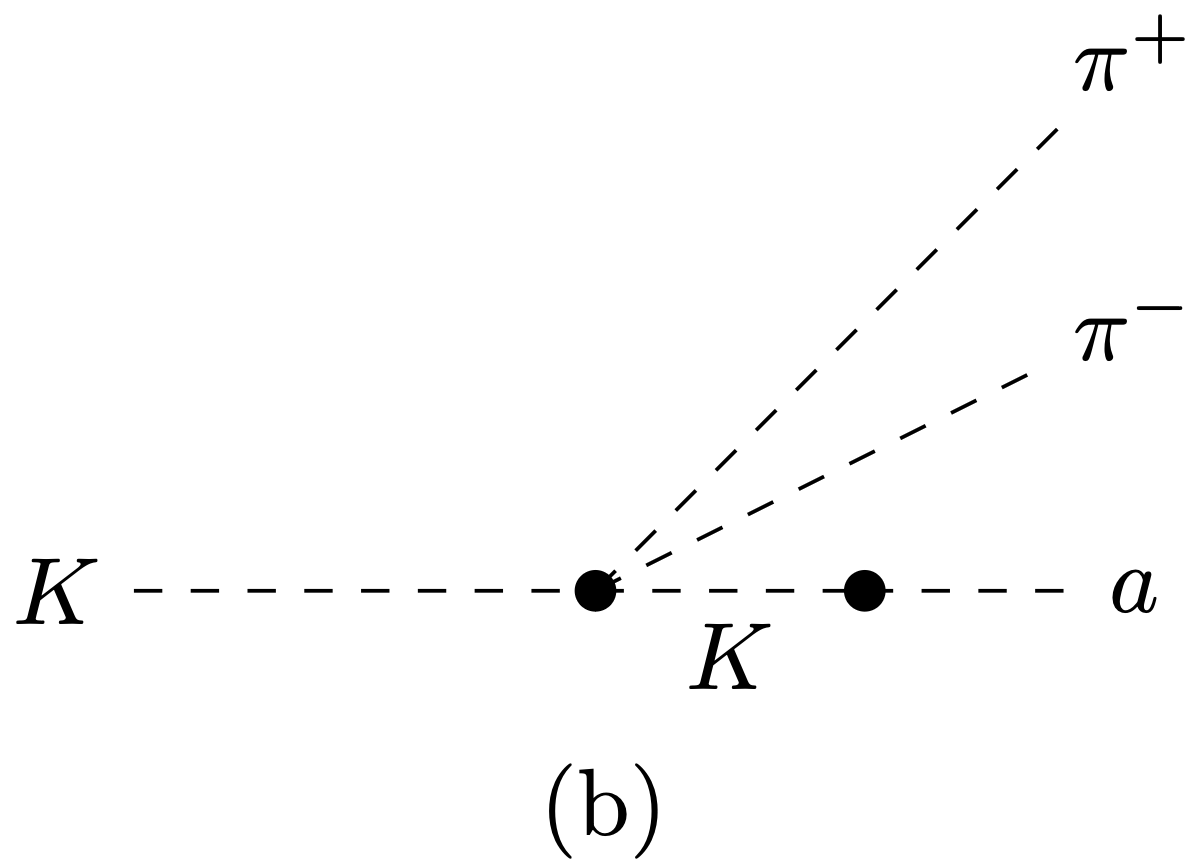}
 \centering
\caption{Tree-level contributions to $K \to \pi \pi a$. The small filled circles denote insertions of $\Lst$.}
\label{fig:K_pipi_a}
\end{figure}
The leading contributions to the $K^0 \to \pi^+ \pi^- a$ amplitude are depicted in \reffig{fig:K_pipi_a}. These contributions read
\bea
\label{eq:M_abc_pipia}
&&\mcM^{(a)}_{\pipi} = \frac{\sqrt{2} (k_A)^*_{23}}{6 F_0 f_a}\left(2 p_a \cdot p_{\pi^{-}}-p_a \cdot p_{\pi^{+}}+p_a \cdot p_{K^0}\right)~, \\ \nn
&&\mcM^{(b)}_{\pipi} = -\frac{\sqrt{2} (k_A)^*_{23}}{12 F_0 f_a} \frac{m_a^2}{m_a^2-m_{K^0}^2} \times \\
&&\hspace{1.25cm}\left(-p_{\pi^{+}} \cdot p_{\pi^{-}}-p_a \cdot p_{\pi^{+}}+2 p_a \cdot p_{\pi^{-}}-2 p_{K^0} \cdot p_{\pi^{+}}+p_{K^0} \cdot p_{\pi^{-}}+p_a \cdot p_{K^0}\right)~.
\eea
Importantly, these tree-level contributions are generated from insertions of $\Lst$, hence they come with no parametric suppression. 
A further interesting property of the $K_S \to \pi^+ \pi^- a$ amplitude, in analogy with the $K_S \to \pi^+ \pi^- \pi^0$ counterpart, is that by $CP$ symmetry it has to be anti-symmetric under the exchange of the two charged pions. As a consequence, to LO in ChPT the $|\Delta S| = 1$ terms in \refeq{LDeltaS1} do {\em not} contribute to the amplitude $\mathcal{M}(K_S \to \pi^0 \pi^+ \pi^-)$, that only receives contributions from the terms transforming as a $\bf 27$ under $SU(3)_L$ and proportional to $g_{27} \simeq 0.3$ \cite{Bijnens:2002vr}. This is the reason why the experimental BR
\be
\BR(K_S \to \pi^+ \pi^- \pi^0) ~=~ (3.5^{+1.1}_{-0.9}) \times 10^{-7}~,
\ee
is as small as $O(10^{-7})$, and so is the error.
Following the argument in the Introduction, detailed further in \refsec{sec:num_res}, this error can be used to estimate the branching-ratio sensitivity of a dedicated search. This sensitivity is then compared to the theoretical calculation of $\BR(K^0 \to \pi^+ \pi^- a)$ to infer the bound on the axial axion-$d$-$s$ coupling.
One can see that the tightness of the bound from this specific mode is the compound result of a small experimental BR and a ``large'' theoretical BR, due to tree-level insertions of $\Lst$.

From the amplitudes \refeq{eq:M_abc_pipia} and their $\bar K^0$ counterparts, one can straightforwardly calculate the branching ratio for $K_S \to \pi^+ \pi^- a$. To this end, we use the relation $\Ks = (K^0 - \bar K^0) / \sqrt2 + O(\bar \epsilon)$ (see e.g. Ref.~\cite{Buras:1998raa}), neglecting terms proportional to $\bar \epsilon = O(10^{-3})$. The $K_S$ amplitude thus reads
\begin{align}
\label{eq:tot_ampl_Ks_pipia}
	&\mc M_{\pi\pi}(K_S)=\dfrac{\re(k_A)_{23}}{2F_0f_a}\dfrac{m_{K^0}^2}{m_{K^0}^2-m_a^2} p_{K_S}\cdot (p_{\pi^-}-p_{\pi^+}) +\\
	&i\dfrac{\im(k_A)_{23}}{6 F_0 f_a}\left[\dfrac{m_{K^0}^2}{m_{K^0}^2-m_a^2}\left(p_{K_S}\cdot (p_{\pi^+}+p_{\pi^-}) + 2 p_{\pi^+}\cdot p_{\pi^-}\right)-\dfrac{2m_{K^0}^2-m_a^2}{ m_{K^0}^2-m_a^2}\left(m_{K^0}^2-m_{\pi^{\pm}}^2 \right)\right]~.\nn
\end{align}
The amplitude has been written as the sum of a term proportional to $\re (k_A)_{23}$, which is antisymmetric under the exchange $\pi^+ \leftrightarrow \pi^-$, plus a term proportional to $\im (k_A)_{23}$, symmetric under this exchange. The amplitude is thus manifestly antisymmetric under $CP$, as expected.

Finally, we note explicitly that the above contributions do not include axion-strahlung radiation, i.e. diagrams in which the axion is emitted by the final-state pions. In fact, a local $\Ks$-$\pi^+$-$\pi^-$ vertex can, to $O(p^2)$ in ChPT, only be obtained through the $|\Delta S| = 1$ terms in \refeq{LDeltaS1}, which is however suppressed by a factor $G_F\,F_0^2 \sim 10^{-7}$ w.r.t. the strong terms in \refeq{Lstrong}.

Formulae for the BR, both in doubly and singly (in $s$) differential form, will be collected in \refsec{sec:BRs}, after discussing the $K \to \ell^+ \ell^- a$ case.

\subsection{\boldmath Application to $K^+ \to \pi^+ \pi^0 a$}\label{sec:Kp}

\noi This process is very similar to $K^0 \to \pi^+ \pi^- a$, apart from the different charges of the initial state and of one of the pions.
The leading-order contributions to the process $K^+ \to \pi^+ \pi^0 a$ are given by
\begin{align}
	\mathcal{M}_{K^+}^{(a)}=&\dfrac{(k_A)_{23}^*}{2F_0f_a}(p_{a}\cdot p_{\pi^0}-p_{a}\cdot p_{\pi^+})~,\\
	\mathcal{M}_{K^+}^{(b)}=&\dfrac{(k_A)_{23}^*}{4F_0f_a}\dfrac{m_a^2}{m_a^2-m_{K^0}^2}(p_{K^+}\cdot p_{\pi^+}-p_{K^+}\cdot p_{\pi^0}+p_{a}\cdot p_{\pi^+}-p_{a}\cdot p_{\pi^0})~.
\end{align}
To $O(p^2)$ in the chiral expansion, the total amplitude reads
\begin{equation}\label{eq:tot_ampl_Kp}
	\mathcal{M}_{K^+}=\dfrac{(k_A)_{23}^*}{2F_0f_a}\dfrac{m_{K^0}^2}{m_{K^0}^2-m_a^2}p_{K^+}\cdot(p_{\pi^0}-p_{\pi^+})~.
\end{equation}
This formula is very similar to the real part of \refeq{eq:tot_ampl_Ks_pipia} but with $\Re(k_A)_{23}\to(k_A)_{23}^*$.

\subsection{\boldmath Application to $K \to \ell^+ \ell^- a$}\label{sec:Klla}

\begin{figure}[b]
 \centering
 \includegraphics[width=0.3\textwidth]{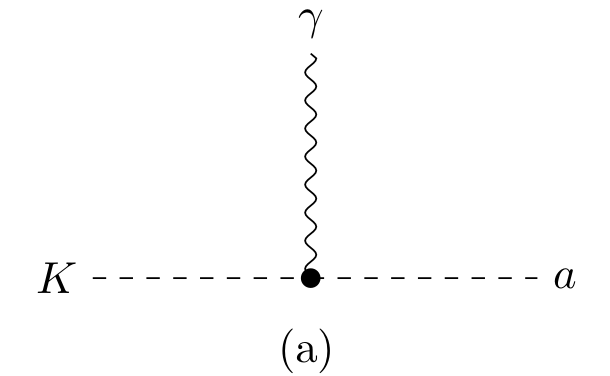}
 \includegraphics[width=0.3\textwidth]{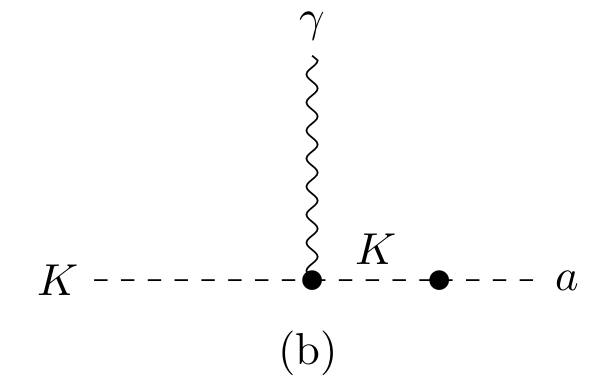}
 \centering
\caption{Tree contributions to $K \to \gamma^* + a$. The small filled circles denote insertions of $\Lst$.}
\label{fig:tree}
\end{figure}

The amplitude for $K \to \ell^+ \ell^- a$ requires a virtual photon or $Z$-boson producing the di-lepton pair. Neglecting the $Z$-boson contribution, the tree-level diagrams are analogous to the $K \to \ell^+ \ell^- \pi^0$ counterpart~\cite{\ecker}. These diagrams are shown in \reffig{fig:tree}.

At $O(p^2)$, vertices with a photon necessarily involve the last term on the r.h.s. of \refeq{eq:DmuU}: since the commutator vanishes if one restricts to the neutral-meson components of $U$, the tree-level contributions to $K \to \ell^+ \ell^- a$ are {\em individually} zero. At $O(p^4)$, vertices with a photon either involve the last term in \refeq{eq:DmuU}, and the same argument applies, or they involve the antisymmetric e.m. tensor. Since the latter has a well-defined classical limit (in which the photon ``sees'' the mesons it interacts with as point-like, uncharged particles), these contributions are also zero individually.
Therefore, in analogy with $K \to \ell^+ \ell^- \pi$, the first non-trivial contributions to the amplitude for $K \to \ell^+ \ell^- a$ arise at one loop with $O(p^2)$ vertices. We next discuss these contributions in detail.

\begin{figure}[t]
 \centering
 \includegraphics[width=0.30\textwidth]{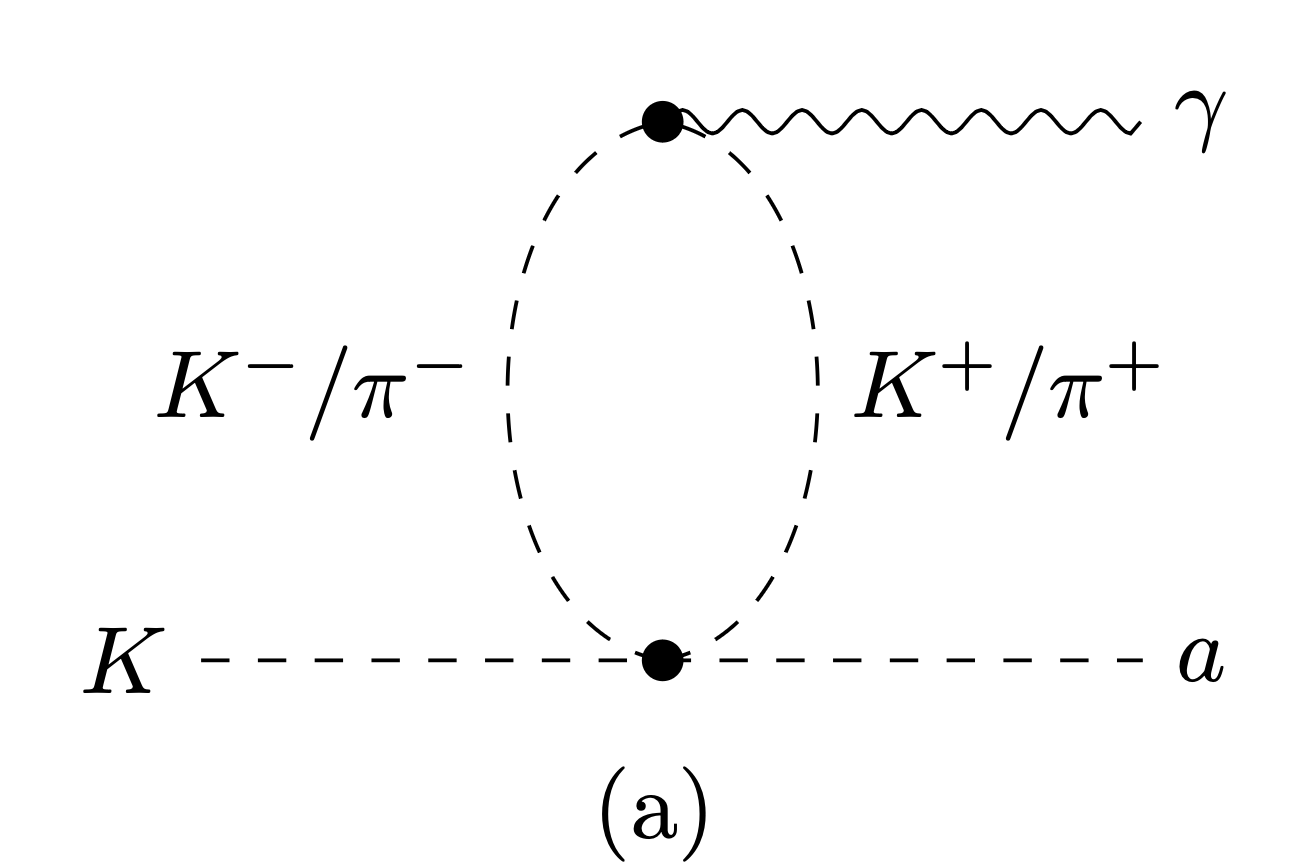}
 \includegraphics[width=0.30\textwidth]{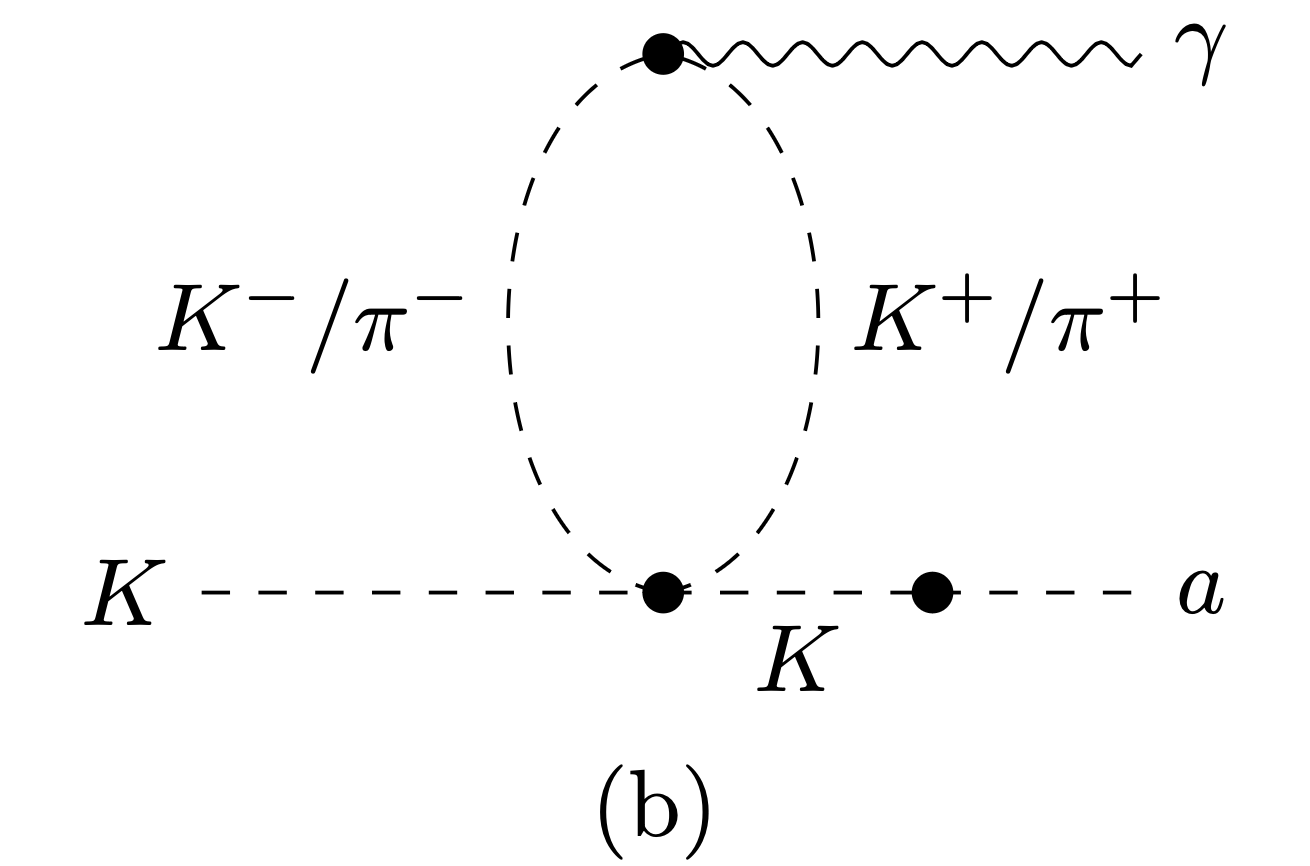}
 \includegraphics[width=0.30\textwidth]{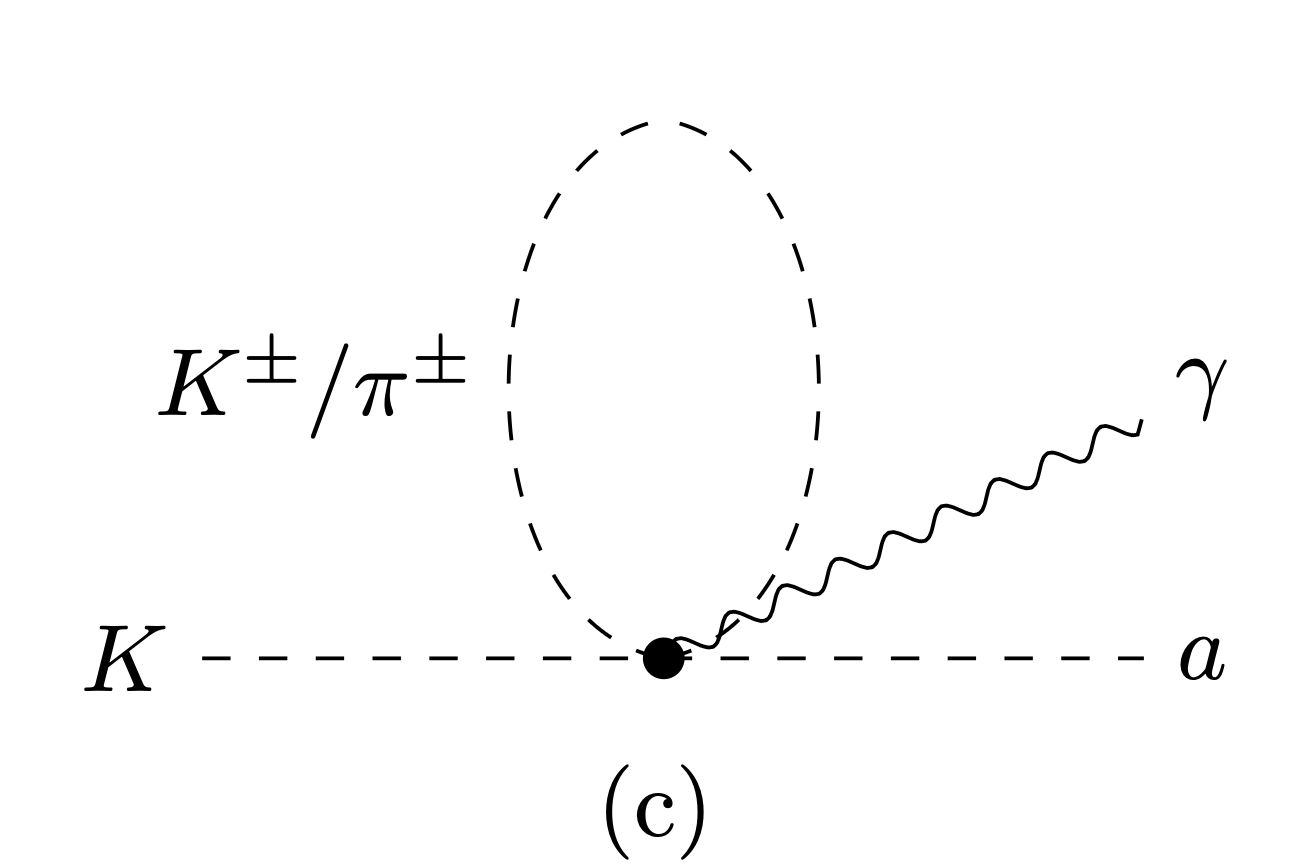}
 \centering
\caption{One-loop contributions to $K \to \gamma^* + a$. The small filled circles denote insertions of $\Lst$.}
\label{fig:1loop}
\end{figure}

Considering only non-weak vertices, i.e. only insertions of $\Lst$ terms, the one-loop diagrams contributing to the process are depicted in \reffig{fig:1loop}.
Similarly as the $K \to \ell^+ \ell^- \pi$ case \cite{\ecker}, the amplitude's structure is constrained by e.m. gauge invariance. Starting from $\mc M \propto \varepsilon^{\mu*} \mc V_\mu (p_K, p_a)$ one must have
\be
\mc V_\mu(p_K, p_a)=\left[p_-^2 (p_+)_\mu - \left(M_{\mathrm{K}}^2-m_a^2\right) (p_-)_\mu\right] \phi\left(p_-^2 / M_{\mathrm{K}}^2, m_a^2 / M_{\mathrm{K}}^2\right)~,
\ee
where $p_{\pm} \equiv p_K \pm p_a$ and $\phi$ is the main function to determine. Since the outgoing momentum of the virtual photon is $q = p_{\ell^+} + p_{\ell^-} = p_K - p_a$, gauge invariance requires that the only contributing terms be the $\mc V_\mu$ components along $p_-^2 (p_+)_\mu = q^2 (p_K + p_a)_\mu$.
We introduce additional definitions similar to Ref.~\cite{\ecker}'s, for the sake of easing comparisons:
\bea
&&I\left(q^2, m_i^2, \mu^2\right)=\frac{1}{16 \pi^2}\left(\frac{1}{\hat{\epsilon}}\left(2 m_i^2-\frac{q^2}{3}\right)+J\left(q^2, m_i^2, \mu^2\right)\right)~,\\
&&\frac{1}{\hat{\epsilon}}=-\frac{1}{\epsilon}+\frac{1}{2}(\gamma_E-1-\ln (4 \pi))+\mathcal{O}(\epsilon)~,
\eea
where $\epsilon = 4 - d$ (with $d$ the dimension used for regularizing the integrals), $\gamma_E$ denotes Euler's constant and $m_i$ denotes the mass in the loop, i.e. either $\MK$ or $\MPI$. This leads to
\be
J\left(q^2, m_i^2, \mu^2\right)~=~\int_0^1 d x~\De_i \ln \left(\frac{\De_i}{\mu^2}\right)~,
\ee
where $\De_i \equiv m^2_{i} - x (1 - x) q^2$. Following Ref.~\cite{\ecker}, in order to project, out of $I$, all and only the components proportional to $q^2 (p_K + p_a)_\mu$, we define the ``subtracted'' $I$ function
\be
\hat{I}\left(q^2, m_i\right)=I\left(q^2, m_i^2\right)-I\left(0, m_i^2\right)~.
\ee
The contribution from the sum of diagrams $(a)$ to the amplitude for $K^0 \to \ell^+ \ell^- a$ thus reads
\bea
\label{eq:Ma}
&&\hspace{-1cm}\mcM^{(a)}_{\mumu} = -i \frac{e^2~(k_A)_{23}^*}{\sqrt2 (4\pi)^2 F_0 f_a q^2} \, \bar u_{\ell^-} \slashed{p}_a v_{\ell^+} I^{(a)} ~,\\
&&\hspace{-1cm}I^{(a)}=
\left(-\frac{2}{\epsilon}+\gamma_E-1\right)\left( \MK^2-\MPI^2\right)
+\int_0^1 \DeK \ln \left(\frac{\DeK}{4 \pi \mu^2}\right) dx
-\int_0^1 \DePI \ln \left(\frac{\DePI}{4 \pi \mu^2}\right) dx~.
\eea
We see that the corresponding subtracted function
\be
\hat I^{(a)} =
\int_0^1 \left( \MK^2 \ln \left(\frac{\DeK}{\MK^2}\right) 
- \MPI^2 \ln \left(\frac{\DePI}{\MPI^2}\right) + \al(1-\al) q^2 \ln \left( \frac{\DePI}{\DeK}\right)
\right) d \alpha
\ee
is {\em finite} and that the cancellation of divergences occurs only in the sum between the $K$- and $\pi$-loop contributions. The corresponding contribution from the sum of diagrams $(b)$ yields
\be
\label{eq:Mb}
\mcM^{(b)}_{\mumu} = -i \frac{e^2~(k_A)_{23}^*}{\sqrt2 (4\pi)^2 F_0 f_a q^2} \, \frac{m_a^2}{m_{K_S}^2 - m_a^2} \bar u_{\ell^-} \slashed{p}_a v_{\ell^+} I^{(b)} ~,
\ee
with $I^{(b)} = I^{(a)}$. Finally, the sum of diagrams $(c)$ yields
\be
\label{eq:Mc}
\mcM^{(c)}_{\mumu}  = -i \frac{e^2~(k_A)_{23}^*}{2 (4\pi)^2 F_0 f_a q^2} \, \bar u_{\ell^-} \slashed{p}_a v_{\ell^+} I^{(c)}~,
\ee
with
\be
I^{(c)} = \left(-\frac{2}{\epsilon}+\gamma_E-1\right)\left( \MK^2-\MPI^2\right) + \MK^2 \ln\left( \frac{\MK^2}{4\pi \mu^2} \right)
- \MPI^2 \ln\left( \frac{\MPI^2}{4\pi \mu^2} \right)~,
\ee
implying no contribution, because $\hat I^{(c)} = 0$.

In short, the NLO result is such that divergences accidentally cancel, due to the interplay between e.m. gauge invariance and chiral symmetry. The important practical consequence is that renormalization-scale dependence is also absent to this order, hence the theory error is only parametric, i.e. tiny.

From the amplitudes Eqs.~(\ref{eq:Ma}), (\ref{eq:Mb}) and (\ref{eq:Mc}) and their $\bar K^0$ counterparts, one can straightforwardly calculate the branching ratio for $K_S \to \ell^+ \ell^- a$, in a way similar as discussed above \refeq{eq:tot_ampl_Ks_pipia}.

\subsection{BR formulae}\label{sec:BRs}

We next discuss our predictions for the doubly  and singly differential, as well as for the total decay widths. The Mandelstam variables are defined as $s = (p_K-p_a)^2$, $t = (p_K-p_{h^+})^2$.\footnote{\label{foot:h} Similarly as in the Introduction, by $h$ we denote the charged pion or lepton, depending on the context.} We also introduce the following abbreviations
\bea
\label{eq:pars}
&&\omega_{\pi} \equiv m_a^2 + m_K^2 + 2 m_{\pi}^2 -s - 2 t, \nn \\
&&\omega_{\ell} \equiv m_a^2 (m_{\ell}^2+m_K^2-t)+(m_{\ell}^2-t)(m_{\ell}^2+m_K^2-s-t), \nn \\
&&\kappa_{\pi} \equiv \frac{s - 4 m_{\pi}^2}{s} \left[m_a^4 + (m_K^2-s)^2 -2 m_a^2(m_K^2+s) \right],\nn \\
&&\kappa_{\ell} \equiv \frac{(s - 4 m_{\ell}^2)^{1/3}}{s} \left[m_a^4 + (m_K^2-s)^2 -2 m_a^2(m_K^2+s) \right],\\
&&\eta_{\pi} \equiv 12288 \pi^3 F_0^2 (m_a^2 - m_K^2)^2,\nn \\
&&\eta_{\ell} \equiv 12288 \pi^5 F_0^2 (m_a^2 - m_K^2)^2 s^2, \nn \\
&&\rho \equiv 3 s -3 m_K^2 + m_a^2 - 2 \dfrac{m_a^2 m_{\pi}^2}{m_K^2}. \nn
\eea
With these definitions, the doubly differential widths read
\bea
&&\frac{d^2\Gamma(K_S \to \pi^+ \pi^- a)}{dsdt} = 3 m_K \frac{\omega_{\pi}^2}{\eta_{\pi}}  \left( \frac{\Re(k_A)_{23}}{f_a} \right)^2 + m_{K} \frac{\rho^2}{3 \eta_{\pi}} \left( \frac{\Im(k_A)_{23}}{f_a} \right)^2~,\nn \\
\label{eq:d2Ga}
&&\frac{d^2\Gamma(K^+ \to \pi^+ \pi^0 a)}{dsdt} = 3 m_K \frac{\omega_{\pi}^2}{\eta_{\pi}}  \left( \frac{|(k_A)_{23}|}{f_a} \right)^2~,\\
&&\frac{d^2\Gamma(K_S \to \ell^+ \ell^- a)}{dsdt} = 6 m_K \frac{\alem^2 \vert\hat I^{(a)}\vert^2 \omega_{\ell}}{\eta_{\ell}} \left( \frac{\Re(k_A)_{23}}{f_a} \right)^2~.\nn
\eea
The $s$ variable is limited to the range  $[4 \,m_{h}^2, (m_K - m_a)^2]$ and $t \in [t_{\min}(s), t_{\max}(s)]$, where the functions $t_{\min,\max}(s)$ are fixed by kinematics \cite{\pdg}. Integrating in $t$ in its kinematical range, we obtain the following singly differential widths
\bea
&&\frac{d\Gamma(K_S \to \pi^+ \pi^- a)}{ds} = \frac{\kappa_{\pi}^{3/2}}{\eta_{\pi}} m_K \left( \frac{\Re(k_A)_{23}}{f_a} \right)^2 + \frac{\kappa_{\pi}^{1/2}\rho^2}{3  \eta_{\pi}} m_K  \left( \frac{\Im(k_A)_{23}}{f_a} \right)^2~, \nn \\
\label{eq:dGa}
&&\frac{d\Gamma(K^+ \to \pi^+ \pi^0 a)}{ds} = \frac{\kappa_{\pi}^{3/2}}{\eta_{\pi}} m_K  \left( \frac{|(k_A)_{23}|}{f_a} \right)^2~,\\
&&\frac{d\Gamma(K_S \to \ell^+ \ell^- a)}{ds} = m_K \frac{\alem^2 \vert\hat I^{(a)}\vert^2 \kappa_{\ell}^{3/2} (2 m_{\ell}^2 + s)}{\eta_{\ell}} \left( \frac{\Re(k_A)_{23}}{f_a} \right)^2~.\nn 
\eea

The dependence on $(k_A)_{23}$ in \refeq{eq:d2Ga} deserves a  comment. The $K_S \to \mumu a$ decay, like $K_S \to \mumu \pi^0$, dominantly  involves the $CP$-even component of the $K_S$, and each of the final states is a $CP$ eigenstate, including the $\mumu$ state, because it originates from a virtual photon. It follows that $\mc M(K_S \to \mumu a) \propto \Re(k_A)_{23}$. In the $K_S \to \pipi a$ case, on the other hand, the $\pipi$ pair has both $CP$-even and -odd components, thus $\mc M(K_S \to \mumu a)$ has both $\Re(k_A)_{23}$ and $\Im(k_A)_{23}$ contributions.

\section{Numerical results} \label{sec:num_res}

\subsection{Total branching ratios}

The most immediate test of the constraining power of our considered decays are the total branching ratios, $\BR(K \to \{\pipi, \mumu \} a)$. We here identify and discuss suitable data to compare against these predictions, and extract the limit on the fundamental couplings that follow from this comparison.

In a dedicated search, the signal for our decays of interest is the total yield (backgrounds subtracted) of $\pipi$ or $\mumu$ pairs with total four momentum $p_{\rm vis}$---assumed to be entirely reconstructed---plus missing four-momentum $\slashed P$, whereby the total four-momentum $(p_{\rm vis} + \slashed P)^2$ is  consistent with $m_K^2$ within a suitable uncertainty. By definition, this {\em total} yield is insensitive to angular anisotropies in the particle(s) that give rise to $\slashed P$, so it is insensitive to the total spin carried by these particles. On the other hand, we can assume that the measurement will in general be sensitive to the invariant mass of the escaping particle(s) through $\sqrt (\slashed P)^2$. Thus, in our case of a single escaping particle, this search will in general be able to differentiate between $m_a \ll m_\pi$ and $m_a = m_\pi$.

Given the described signal, we identify the main irreducible backgrounds depending on the external hadrons ($K_S \to \pippim$, $K^+ \to \pi^+ \pi^0$, $K_S \to \mupmum$) and on the ALP mass.
Before providing details on each channel, we note here some general features.
The expected yield from each identified background has two main sources of uncertainty: the absolute uncertainty on the given experimental branching ratio, denoted as $\sigma_{\rm bkg}$, and the uncertainty on the yield of the normalization channel. In our study, the former is taken from the PDG~\cite{\pdg}, while the latter is assumed to have a systematic component at the level of 5\% of the central value of the background BR, denoted as $\BR_{\rm bkg}$.\footnote{The 5\% value is an educated, though otherwise arbitrary, guess. This serves to safeguard our $\Seff$ definition \refeq{eq:Seff} from cases where the error $\sigma_{\rm bkg}$ is very small, so that \refeq{eq:Seff}, without the $0.05 \times \BR_{\rm bkg}$ component, would produce an overly optimistic sensitivity estimate. Examples of analyses supporting our guess include e.g. Ref.~\cite{LHCb:2022tpr}, see its Table I. We also note that sources of systematics related to the hardware and high-level triggers will be smaller or absent in future measurements.}
To the extent that the signal and the irreducible-background yields are normalized to the same channel, the quantity 
\be
\label{eq:Seff}
\Seff \equiv \sqrt{(0.05 \times \BR_{\rm bkg})^2 + \sigma_{\rm bkg}^2}
\ee
seems to realistically quantify the best sensitivity attainable in a dedicated search. In simple terms, the sensitivity to the signal BR may be estimated from the uncertainty on the leading background(s) present in the search. Put differently, the signal can at most be as large as the extent to which its dominant sources of irreducible background are controlled. We emphasize that $\Seff$ is a sensitivity, not a physical branching-ratio limit, that only a dedicated search can determine. However, $\Seff$ does quantify how well (in terms of BR limit) such search can possibly do, based on how well the leading irreducible backgrounds are controlled.

With this important qualification in mind, we next use $\Seff$ to place bounds on the theoretically calculated BR, to be denoted as $\BRth$. These bounds will be obtained from relations of the form $\BRth((k_{A})_{23}) = \Seff$, and will show that dedicated searches in each of our considered channels would provide novel and very effective probes of the considered couplings. We will show that this is true even for $f_a$ as large as $10^6~\TeV$ and $m_a \ll m_\pi$, namely for ALP masses
approaching the parameter space of the QCD axion.

To be self-contained, we first collect the values of the measured BRs relevant for the $\Seff$ calculation in our $K \to \{\pipi, \mumu\} a$ channels of interest, assuming $m_a$ is either negligible with respect to the external states' or, respectively, equal to the pion mass. From Ref.~\cite{\pdg} one finds
\be
	\label{eq:BRs_exp}
\hspace{-0.2cm}\begin{array}{llll}
		\BR(K_S \to \pi^+ \pi^- \gamma) &=~ (1.79 \pm 0.05) \times 10^{-3}~, ~~ 	&\BR(K_S \to \pi^+ \pi^- \pi^0) &=~ (3.5^{+1.1}_{-0.9}) \times 10^{-7}~, \\
		[0.3cm]
		\BR(K_L \to \pi^+ \pi^- \gamma) &=~ (4.15 \pm 0.15) \times 10^{-5}~,~~ 	&\BR(K_L \to \pi^+ \pi^- \pi^0) &=~ (12.54 \pm 0.05) \times 10^{-2}~, \\
		[0.3cm] 
		\BR(K_S \to \pi \mu \nu) &=~ (4.56 \pm 0.20) \times 10^{-4}~,~~ &\BR(K_S \to \mu^+ \mu^- \pi^0) &=~ (2.9^{+1.5}_{-1.2}) \times 10^{-9}~, \\
		[0.3cm]
		\BR(K_L \to \pi \mu \nu) &=~ (27.04 \pm 0.07) \times 10^{-2}~,~~ &\BR(K^+ \to \pi^+ \pi^0 \pi^0) &=~ (1.760 \pm 0.023) \times 10^{-2}~, \\
		[0.3cm]
		\BR(K^+ \to \pi^+ \pi^0 \gamma) &=~ (6.0 \pm 0.4) \times 10^{-6}~, ~~&\BR(K^+ \to \mu^+ \pi^0 \nu) &=~ (3.352 \pm 0.034) \times 10^{-2}~,
	\end{array}
\ee
whereas $\BR(K_S \to \mu^+ \mu^- \gamma)$ is unmeasured.
Using \refeq{eq:Seff} on \refeq{eq:BRs_exp}, we can then stipulate the following approximate relations
\newcommand{\shiftl}{\hspace{-1cm}}
\be
\label{eq:Seff_num}
\begin{array}{lll}
\BRth(K_S \to \pi^+ \pi^- a)_{m_a \ll m_\pi} & \lesssim~~ \Seff(K_S \to \pippim \gamma) & \shiftl \approx~~ 1.0 \times 10^{-4}~, \\ [0.35cm]
\BRth(K_S \to \pi^+ \pi^- a)_{m_a = m_\pi} & \lesssim~~ \Seff(K_L \to \pippim \pi^0, \AKL) & \shiftl \approx~~ 1.3 \times 10^{-5}~, \\ [0.35cm]
\BRth(K^+ \to \pi^+ \pi^0 a)_{m_a \ll m_\pi} & \lesssim~~ \Seff(K^+ \to \pi^+ \pi^0 \gamma) & \shiftl \approx~~ 5.0 \times 10^{-7}~, \\ [0.35cm]
\BRth(K^+ \to \pi^+ \pi^0 a)_{m_a = m_\pi} & \lesssim~~ \Seff(K^+ \to \pi^+ \pi^0 \pi^0) & \shiftl \approx~~ 9.1 \times 10^{-4}~~, \\ [0.35cm]
\BRth(K_S \to \mu^+ \mu^- a)_{m_a \ll m_\pi} & \lesssim~~ \Seff(K_{S,L} \to \pi \mu \nu, K_S \to \pipi \gamma, \epimu,\AKL) & \shiftl \approx ~~ 4.1 \times 10^{-7}, \\ [0.35cm]
\BRth(K_S \to \mu^+ \mu^- a)_{m_a = m_\pi} & \lesssim~~ \Seff(K_{S} \to \mupmum \pi^0,K_{L} \to \pippim \pi^0,\epimu,\AKL) & \\
 & & \shiftl \approx~~ 2.0 \times 10^{-9}~.
\end{array}
\ee
\newcommand{\emupiLHCb}{\emupi}
A few comments on the above relations follow. As regards the $K_S \to \pippim a$ channel with small $m_a$ (first in \refeq{eq:Seff_num}), a first background is given by $K_S \to \pippim \gamma$ with a hard yet undetected $\gamma$. Using \refeq{eq:BRs_exp} and \refeq{eq:Seff}, one finds $\Seff \approx 1 \times 10^{-4}$. A further background is represented by $K_{S} \to \pi \mu \nu$ (charge-conjugated modes included in the final state), with one muon misidentified as a pion. Quantifying the $\mu$-to-$\pi$ mis-identification probability at LHCb as $\emupiLHCb = 1 \times 10^{-2}$ \cite{LHCb:2014set}, this channel yields an $\Seff$ of order $10^{-5} \times \emupiLHCb$, thus subdominant w.r.t. $K_S \to \pippim \gamma$. Two additional irreducible backgrounds would be represented by the $K_L$ counterparts of the above decays.
The corresponding yields are suppressed by the small $K_L$ acceptance within LHCb, which is consequence of the $K_L$ long lifetime relative to the $K_S$'s. We will denote this acceptance factor as $\AKL$. One finds $\AKL \approx 2 \times 10^{-3}$ \cite{\alves}.
On the other hand, the $K_L \to \pi \mu \nu$ channel has a very large BR, see \refeq{eq:BRs_exp}. Including all factors thus gives $\Seff \approx 1.4 \times 10^{-2} \times \emupiLHCb \times \AKL$, which is numerically negligible. A similar conclusion holds for the $K_L \to \pippim \gamma$ channel, whose BR is per se very small. These considerations justify our final $\Seff \approx 1 \times 10^{-4}$ in this channel, see \refeq{eq:Seff_num}.

Following the same logic, we next comment on the $K_S \to \pippim a$ channel with $m_a = m_\pi$. The irreducible background constituted by $K_S \to \pippim \pi^0$ yields $\Seff \approx 1.1 \times 10^{-7}$. Its $K_L$ counterpart yields $\Seff \approx 1.3 \times 10^{-5}$, which includes the acceptance suppression by $\AKL$. We quote the latter as the sensitivity in this channel.

For the $K^+ \to \pi^+ \pi^0 a$ channel with small $m_a$, the most straightforward background comes from $K^+ \to \pi^+ \pi^0 \gamma$, that yields $\Seff \approx 5.0 \times 10^{-7}$. Additionally, the channel $K^+ \to \mu^+ \pi^0 \nu$, with a misidentified muon, comes with a large BR. However, the $\mu$-to-$\pi$ mis-ID probability at NA62\footnote{We thank Cristina Lazzeroni and Joel Swallow for exchanges on this point.} is as small as $1.3 \times 10^{-8}$, making this channel entirely negligible.

As regards the $K_S \to \mupmum a$ channel with $m_a = m_\pi$, the $K_{S} \to \mupmum \pi^0$ and the $K_{L} \to \pippim \pi^0$ backgrounds contribute respectively $\Seff \approx \{1.5, 1.3\} \times 10^{-9}$, whereby the latter takes also into account an $\epimu^2 \times \AKL$ factor, where $\epimu$ denotes the $\pi \to \mu$ mis-ID probability at LHCb, evaluated to 1\%. Adding these two $\Seff$ values in quadrature, we arrive at the quoted $\Seff \approx 2.0 \times 10^{-9}$.

A final comment concerns $\BR(K_S \to \mu^+ \mu^- a)$ with small $m_a$ (penultimate in \refeq{eq:Seff_num}). As noted below \refeq{eq:BRs_exp}, $\BR(K_S \to \mu^+ \mu^- \gamma)$ has not been measured yet. However, several existing channels seem to be usable to the same effect. First, $\BR(K_S \to \pi \mu \nu)$ again: taking into account the $\pi \to \mu$ mis-ID probability at LHCb, $\epimu$, one obtains $\Seff(K_S \to \pi \mu \nu) \simeq 3 \times 10^{-7}$. A second channel is $K_S \to \pipi \gamma$ (first in \refeq{eq:BRs_exp}) with two $\epimu$ factors, which gives $1 \times 10^{-8}$, thus negligible w.r.t. to the former channel. Finally, a third channel is $K_L \to \pi \mu \nu$. Taking into account the $\epimu \times \AKL$ factor yields $\Seff \approx 2.7 \times 10^{-7}$ in this channel. In conclusion, adding the two larger channels in quadrature, one obtains $\Seff \approx 4.1 \times 10^{-7}$, as in the last line of \refeq{eq:Seff_num}.

Interestingly, the decay $K_S \to \mupmum \pi^0$ can be measured by LHCb, both with
and without an explicit $\pi^0$ requirement. In order to separate it from $K_S \to \mupmum a$, with $m_a = m_\pi$, we can use the measurement from NA62 along the above lines and, in the long term, a hypothetical measurement by LHCb with explicit $\pi^0$ requirement. Assuming that the latter reaches a 8\% precision~\cite{Chobanova:2016laz} the effective BR in the last line of \refeq{eq:Seff_num} would be around $1.5 \times 10^{-10}$, i.e. improve by an order of magnitude w.r.t. the $\Seff$ value quoted in \refeq{eq:Seff_num}.

\begin{figure}[t]
 \centering
 \includegraphics[width=0.47\textwidth]{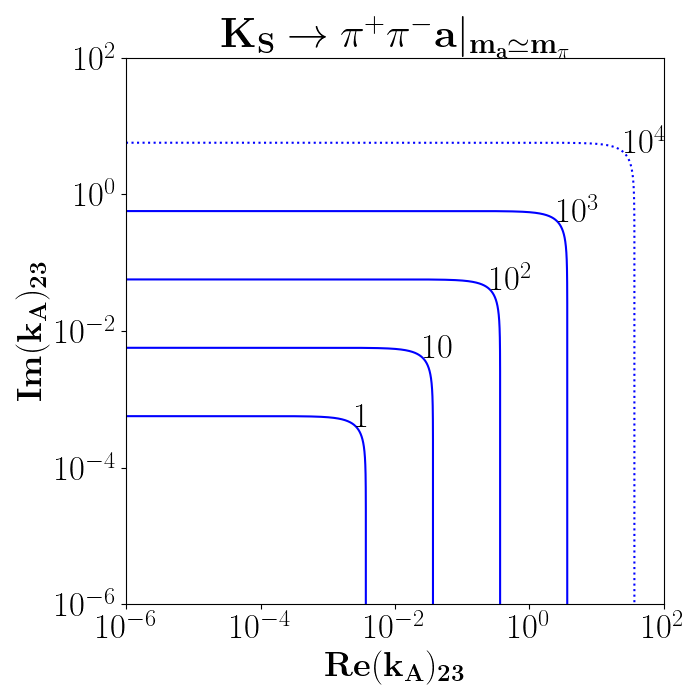} \hfill
 \includegraphics[width=0.47\textwidth]{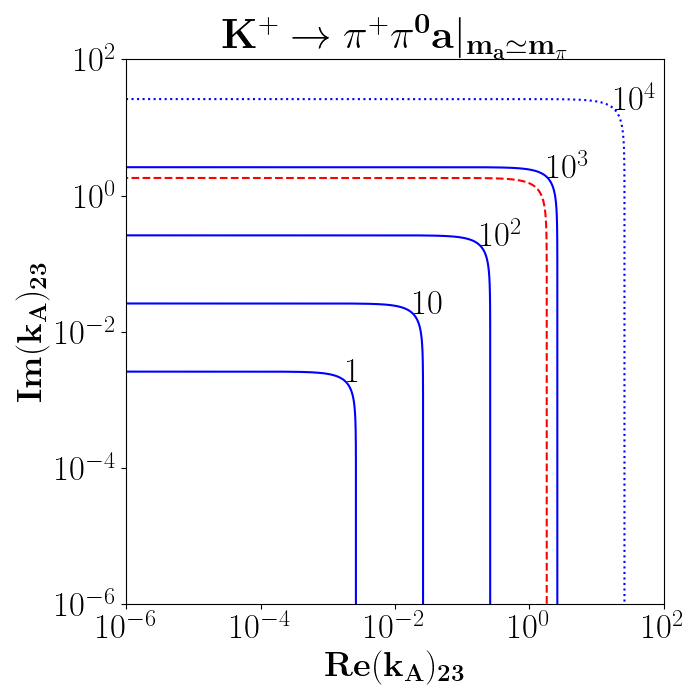}
 \includegraphics[width=0.47\textwidth]{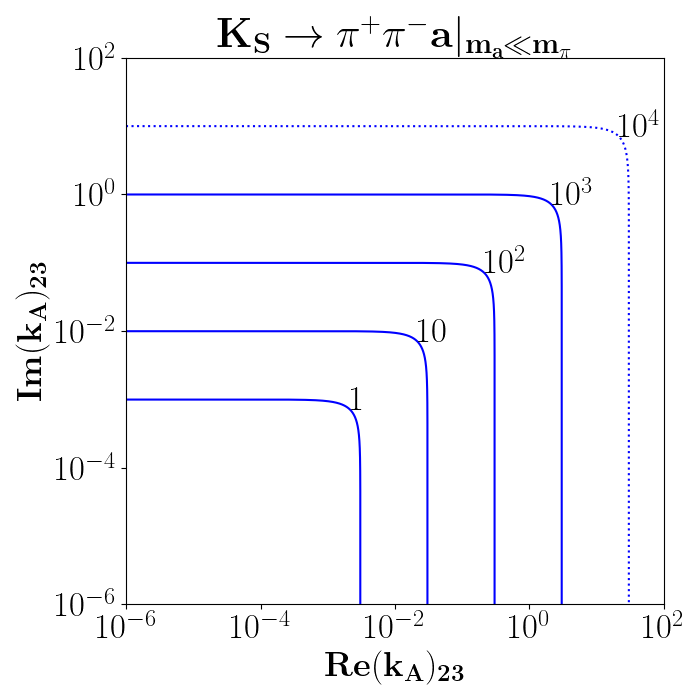} \hfill
 \includegraphics[width=0.47\textwidth]{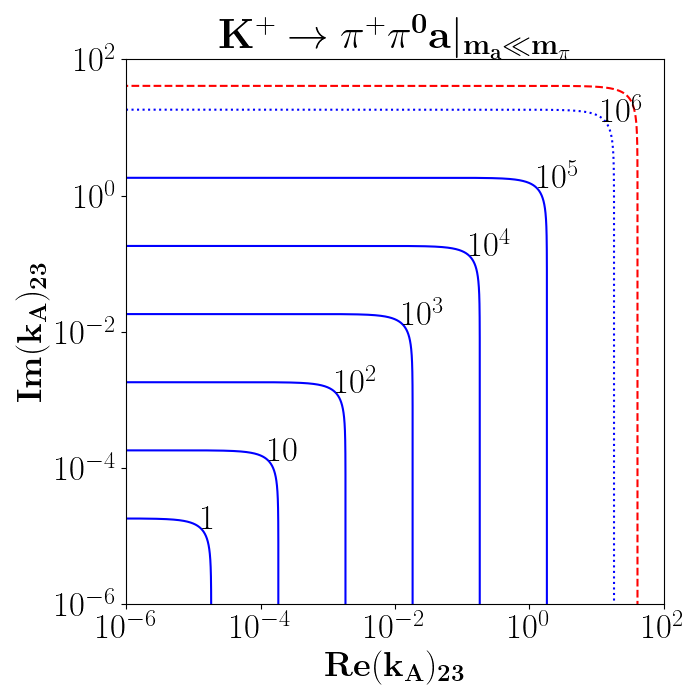}
 \centering
\caption{Lines of constant $f_a$, in TeV, in the plane $\re$ vs. $\im(k_A)_{23}$.
The upper panels compare $K_S \to \pippim a$ vs. $K^+ \to \pi^+ \pi^0 a$ for $m_a = m_\pi$. The lower panels show the same comparison for $m_a \ll m_\pi$.}
\label{fig:ReImkA23_from_BRs}
\end{figure}
The inequalities in \refeq{eq:Seff_num} can be used to estimate the sensitivity to the ratio $(k_A)_{23} / f_a$ from the different channels. ``Saturating'' the relations in \refeq{eq:Seff_num}, i.e. treating them as equalities, one can infer the maximal value of that ratio. The values are collected, channel by channel, in \reftab{tab:kA23overfa}.
\begin{table}[h!]
\begin{centering}
\renewcommand{\arraystretch}{1.4}
\captionsetup{justification=centering, font=small}
\setlength{\tabcolsep}{8pt}

\begin{tabular}{>{\centering\arraybackslash}p{2.5cm}
                >{\centering\arraybackslash}p{1.5cm} 
                >{\centering\arraybackslash}p{1.5cm} 
                >{\centering\arraybackslash}p{1.5cm}| 
                >{\centering\arraybackslash}p{1.5cm} 
                >{\centering\arraybackslash}p{1.5cm} 
                >{\centering\arraybackslash}p{1.5cm}}
\toprule
\rowcolor{lightgray!20}
\multicolumn{1}{c}{} & \multicolumn{3}{c|}{$m_a\ll m_{\pi}$} & \multicolumn{3}{c}{$m_a= m_{\pi}$} \\
\midrule
\rowcolor{lightgray!10}
\textbf{Channel} & $\frac{|\Re[(k_A)_{23}]|}{f_a}$ & $\frac{|\Im[(k_A)_{23}]|}{f_a}$ & $\frac{|(k_A)_{23}|}{f_a}$ & $\frac{|\Re[(k_A)_{23}]|}{f_a}$ & $\frac{|\Im[(k_A)_{23}]|}{f_a}$ & $\frac{|(k_A)_{23}|}{f_a}$ \\
\midrule
$K_S\to\mu^+\mu^-a$ & $48$ & --- & --- & $5.7$ & --- & --- \\
\rowcolor{lightgray!5}
$K_S\to\pi^+\pi^-a$ & $3.0$ & $1.0$ & --- & $3.7$ & $0.57$ & --- \\
$K^+\to\pi^+\pi^0a$ & --- & --- & $0.018$ & --- & --- & $2.6$ \\
\bottomrule
\end{tabular}
\end{centering}
\caption{
Constraints on the ratio of coupling to $f_a$, in units of $10^{-3} \, \mathrm{TeV}^{-1}$, from \refeq{eq:Seff_num}. A separate constraint is provided for each of the channels considered (table rows), distinguishing cases of small versus large $m_a$ as well as constraints on $\re (k_A)_{23}$, $\im (k_A)_{23}$, or $|(k_A)_{23}|$ where applicable (table columns). See the text for more details.
}
\label{tab:kA23overfa} 
\end{table}
With the coupling-over-$f_a$ ratio fixed in this way channel by channel, one can determine the $f_a$ value as a function of the coupling value assumed. This relation is shown in \reffig{fig:ReImkA23_from_BRs} in the plane $\Re(k_A)_{23}$ vs. $\Im(k_A)_{23}$. The coupling and $f_a$ increases, of course, proportionally to each other. Identifying a reference, but otherwise arbitrary, value for a ``large'' coupling allows to determine the maximal $f_a$ scale probed by each channel. We determine this scale as the $f_a$ value for which the real or imaginary part of $(k_A)_{23}$, or its absolute value depending on the channel, is equal (or the closest) to 10. The contour with the largest $f_a$ value in each panel of \reffig{fig:ReImkA23_from_BRs} corresponds to this ``large'' coupling choice.

Our sensitivity study motivates the pursuit of dedicated analyses. It is interesting to compare our results with the available experimental searches, that to our knowledge concern $K^+ \to \pi^+ \pi^0 a$ only.
The most recent dedicated search is presented in Ref.~\cite{Sadovsky:2023cxu} for various choices of the mass of the ALP particle.\footnote{The previous best upper limit could be extracted from Ref.~\cite{Tchikilev:2003ai}, whereas the PDG quotes Ref.~\cite{\adler}. Both these searches use a pure phase-space hypothesis, which Ref.~\cite{Sadovsky:2023cxu} shows to be a crude approximation.} In the case $m_a \ll m_\pi$, the bound inferred on $|(k_A)_{23}|$ from the relevant line in \refeq{eq:Seff_num} is only a factor of $\sim 2$ stronger than the bound obtained through Ref.~\cite{Sadovsky:2023cxu}.
In the case of $m_a = m_\pi$, the bound inferred from the relevant relation in \refeq{eq:Seff_num} is still compatible with Ref.~\cite{Sadovsky:2023cxu}'s, but $\sim 10$ weaker.
The comparison is shown visually in the right-hand side panels of \reffig{fig:ReImkA23_from_BRs}, which pertain to the $K^+$ channel. The red dashed lines are obtained by replacing $\Seff$ in \refeq{eq:Seff_num} with the actual BR limit from Ref.~\cite{Sadovsky:2023cxu}, depending on the $m_a$ value. Additionally, the $f_a$ value is set equal to that in the outermost contour (indicated by a dotted line) in each respective panel. In other words, the red dashed lines are actual limits, obtained by identifying a BR calculation in ChPT (with no a priori assumption on the axion mass) with the experimental bound in Ref.~\cite{Sadovsky:2023cxu}. These limits can be rewritten as
\be
\label{eq:bound_Kp}
\frac{|(k_A)_{23}|}{f_a} \le 10^{-3}~\mathrm{TeV}^{-1}\times \left\{ \begin{array}{ll} 0.041 & \text{for $m_a \ll m_\pi$} \\ 0.18 & \text{for $m_a = m_\pi$} \end{array} \right.~.
\ee

These findings suggest that our simple procedure produces quite consistent results with an actual search in all cases available for such a comparison, namely the $K^+ \to \pi^+ \pi^0 a$ modes. By the same token, our results suggest that dedicated searches in the $K_S$ channels we suggest may even provide order-of-magnitude improvements on the $(k_A)_{23}$ constraint.

\subsection{\boldmath Branching ratios differential in $s$}

\begin{figure}[h!]
	\centering
	\includegraphics[width=0.75\textwidth]{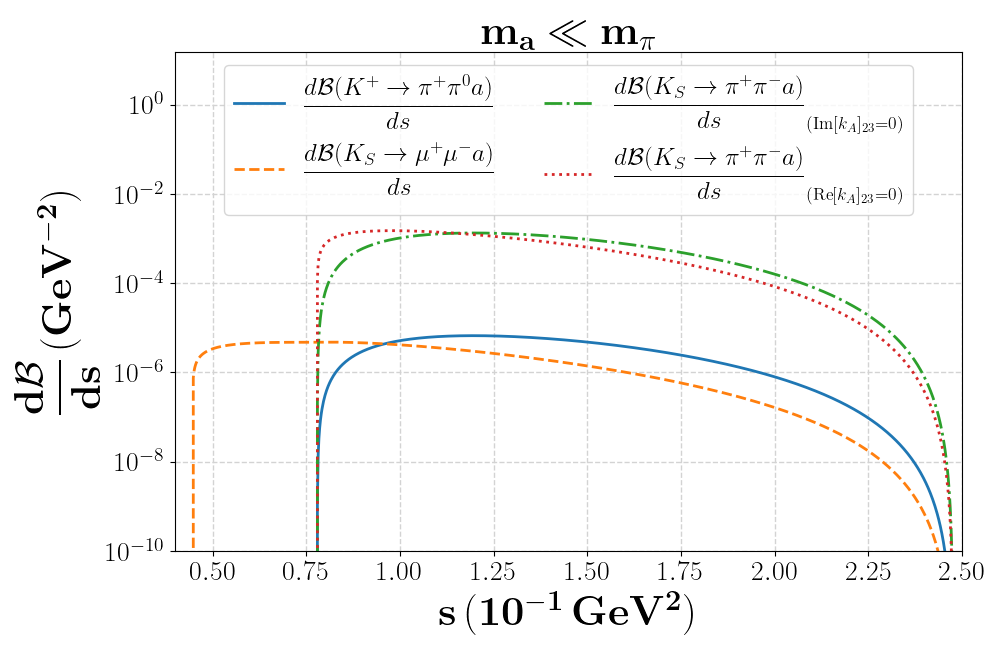}
	\includegraphics[width=0.75\textwidth]{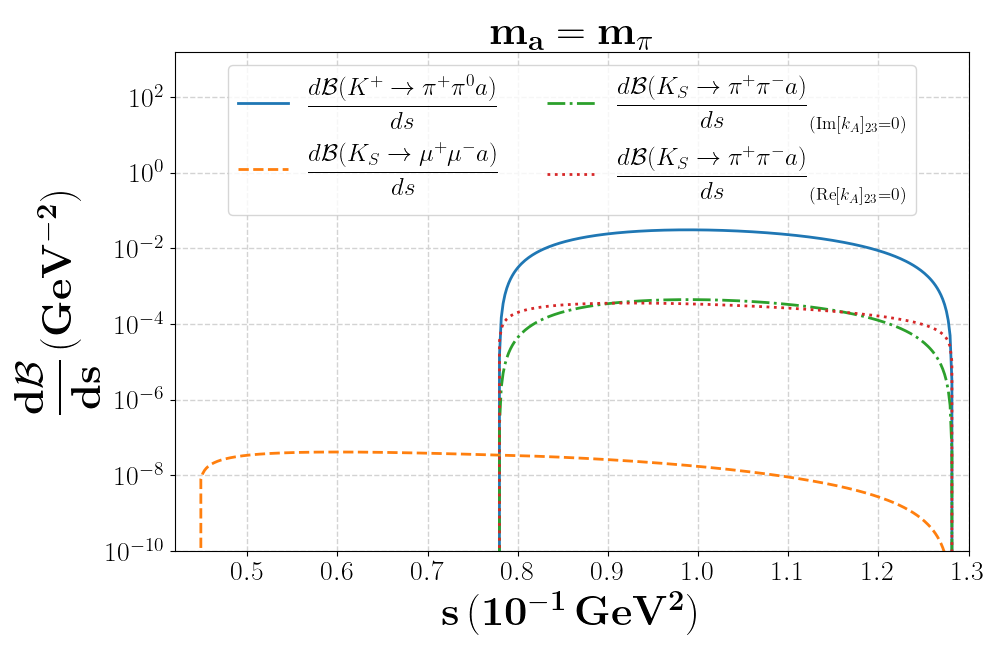}
	\centering
	\caption{
	Singly differential branching ratios for the different $K \to \{\pipi, \mumu \}a$ channels considered. The two panels correspond to $m_a \ll m_\pi$ (upper) or $m_a = m_\pi$ (lower). The couplings are set by assuming saturation of the bounds in \refeq{eq:Seff}, corresponding to a ``maximal'' coupling value, separately for each channel, as discussed around \reftab{tab:kA23overfa}. 
	}
	\label{fig:dBR_ds}
\end{figure}

In each of the channels shown in \reffig{fig:ReImkA23_from_BRs} we can extract the ``maximal'' value of the ratio coupling-over-$f_a$, namely the ratio that saturates the corresponding bound in \refeq{eq:Seff_num}. These ratio values are collected in \reftab{tab:kA23overfa}. We opted for deriving a separate ratio value for each considered channel, further separating the cases of small or large $m_a$. By inspection of \refeq{eq:dGa}, only the $K_S \to \pippim a$ depends separately on $\re(k_A)_{23}$ and $\im(k_A)_{23}$. In this case, we consider different hypotheses on the short-distance coupling, as shown in the table. The constraint on the real (imaginary) part of $(k_A)_{23}$ is obtained assuming that the imaginary (real) part is zero. For each bound, we keep only the first two significant digits. We use the ratios in \reftab{tab:kA23overfa} to determine the ``maximal'' spectra predicted for the singly differential branching ratios. The latter are shown in \reffig{fig:dBR_ds}.

\subsection{Doubly differential widths}

The bound one can infer from a total BR averages out regions with smaller and larger sensitivity to produce an overall bound. In this section, we would like to consider the possibility to improve the bound inferred from a total BR with one derived by considering only the regions in the Dalitz plot which have concurrently the largest yield and the largest reconstruction efficiency.

\begin{figure}[h!]
 \centering
 \includegraphics[width=0.49\textwidth]{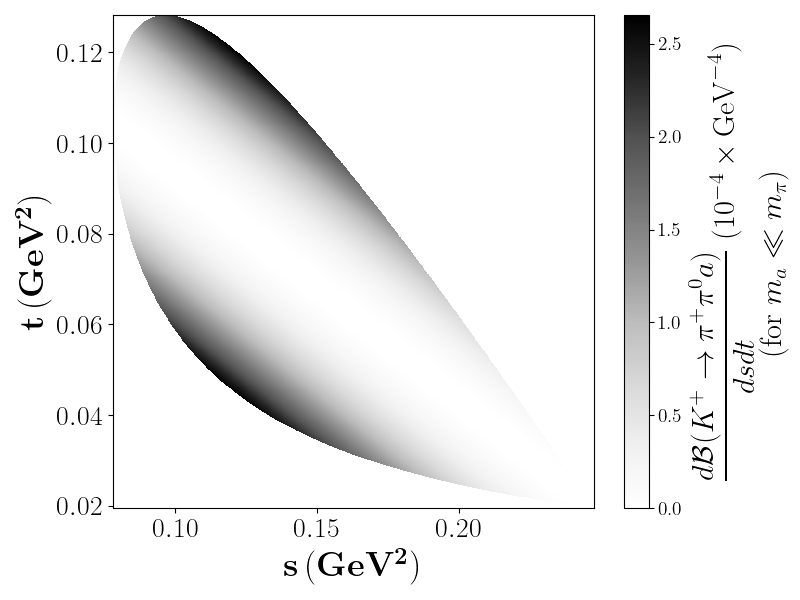}
 \includegraphics[width=0.49\textwidth]{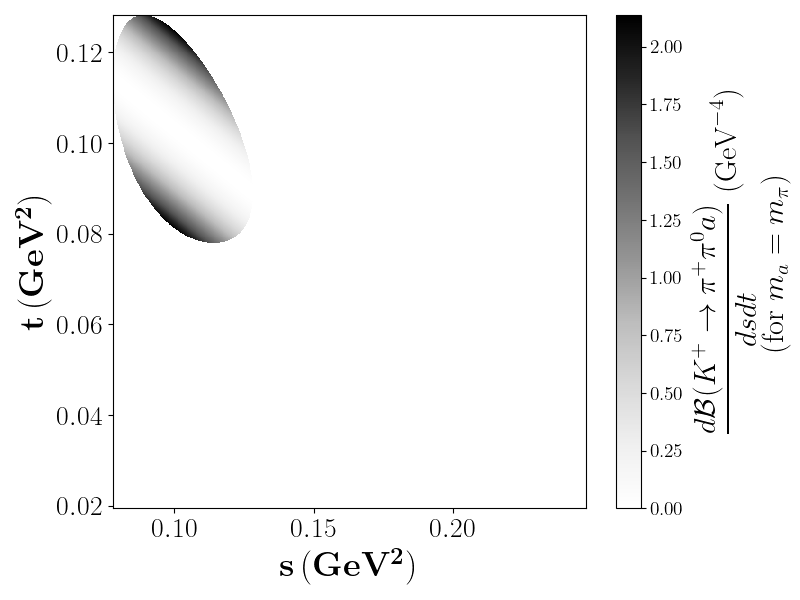}
 \centering
\caption{Doubly differential $\BR(K^+ \to \pi^+\pi^0 a)$ in the $\{s,\,t\}$ plane, with $m_a \ll m_\pi$ (left) and $m_a = m_\pi$~MeV (right). In either case we assume the axion coupling strength to saturate the relevant inequality in \refeq{eq:Seff}.}
\label{fig:dBR_dsdt_Kp}
\end{figure}

The doubly differential BRs for our considered channels are shown in the $s$ vs. $t$ plane in \reffigs{fig:dBR_dsdt_Kp}{fig:dBR_dsdt_KS_mumu}. The four figures show the following cases: $K^+ \to \pi^+\pi^0 a$, $K_S \to \pi^+\pi^- a$ from $\re(k_A)_{23}$ or from $\im(k_A)_{23}$, and $K_S \to \mu^+\mu^- a$, respectively. The two panels in each figure refer to the $m_a \ll m_\pi$ and $m_a = m_\pi$ instances, respectively. 

\reffigs{fig:dBR_dsdt_Kp}{fig:dBR_dsdt_KS_pipi_real} show that the Dalitz plot is completely dominated by two relatively narrow slivers of phase space. By inspection of \refeq{eq:dGa} we see that the BRs in the concerned channels depend on $s,t$ through $\omega_\pi^2$. The vast region in the Dalitz with small local BR thus corresponds to small $\omega_\pi$, i.e. $s + 2t \approx m_a^2 + m_K^2 + 2 m_{\pi}^2$, whereas the two slivers with most of the statistics occur for maximal or minimal $s + 2t$.

\begin{figure}[h!]
\centering
\includegraphics[width=0.49\textwidth]{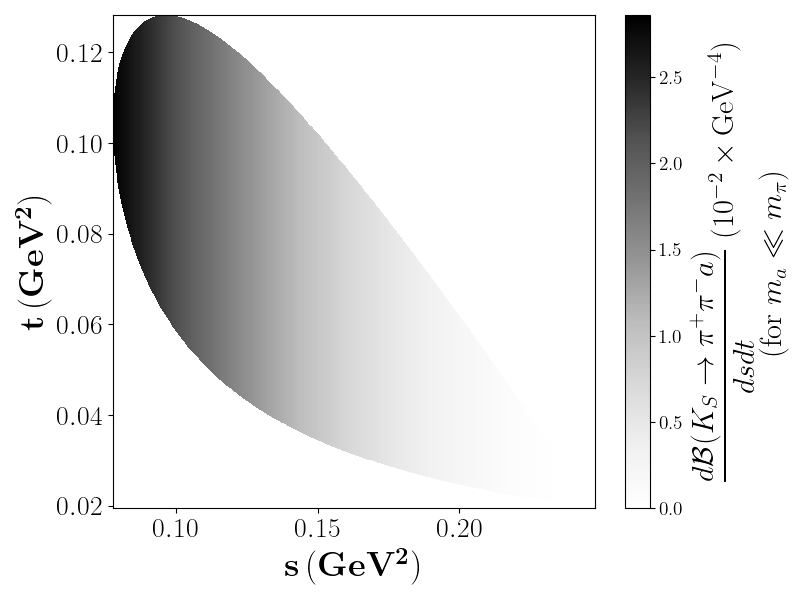}
\includegraphics[width=0.49\textwidth]{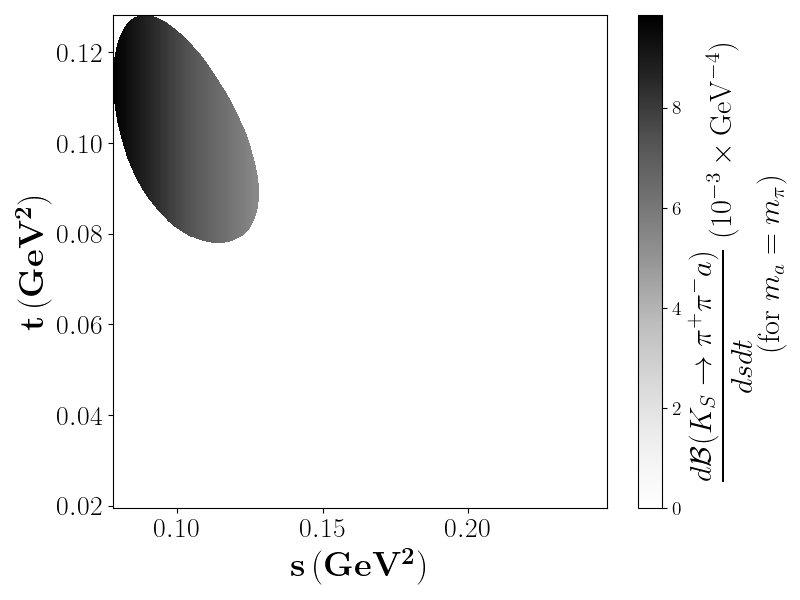}
\centering
\caption{Same as \reffig{fig:dBR_dsdt_Kp}, but for the contribution to $\BR(K_S \to \pi^+\pi^- a)$ proportional to $\im(k_A)_{23}$. 
	}
\label{fig:dBR_dsdt_KS_pipi_imag}
\end{figure}

On the other hand, \reffigs{fig:dBR_dsdt_KS_pipi_imag}{fig:dBR_dsdt_KS_mumu} suggest the region of lower $s$ to be the most important, irrespective of the $t$ value, but this preference is much less pronounced than in \reffigs{fig:dBR_dsdt_Kp}{fig:dBR_dsdt_KS_pipi_real}.
\newcommand{\fastsim}{Chobanova:2020vmx}
 
\begin{figure}[h!]
\centering
\includegraphics[width=0.49\textwidth]{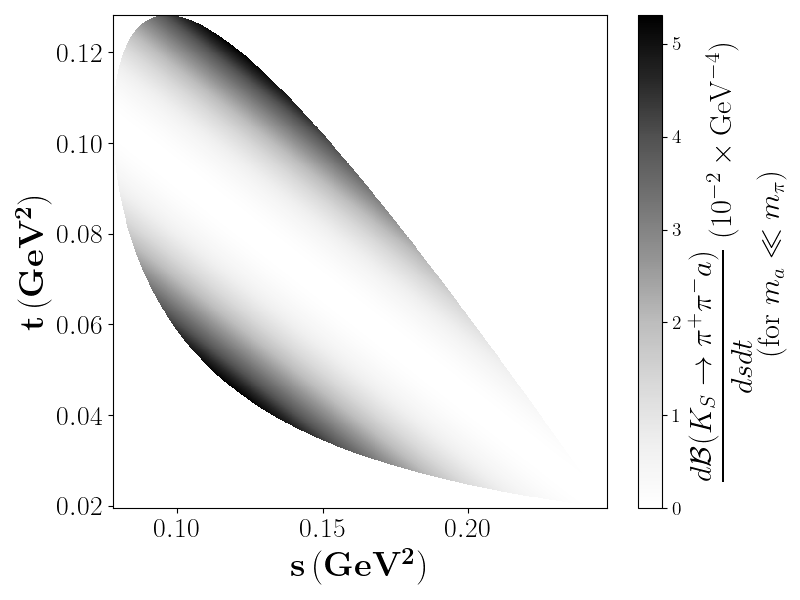}
\includegraphics[width=0.49\textwidth]{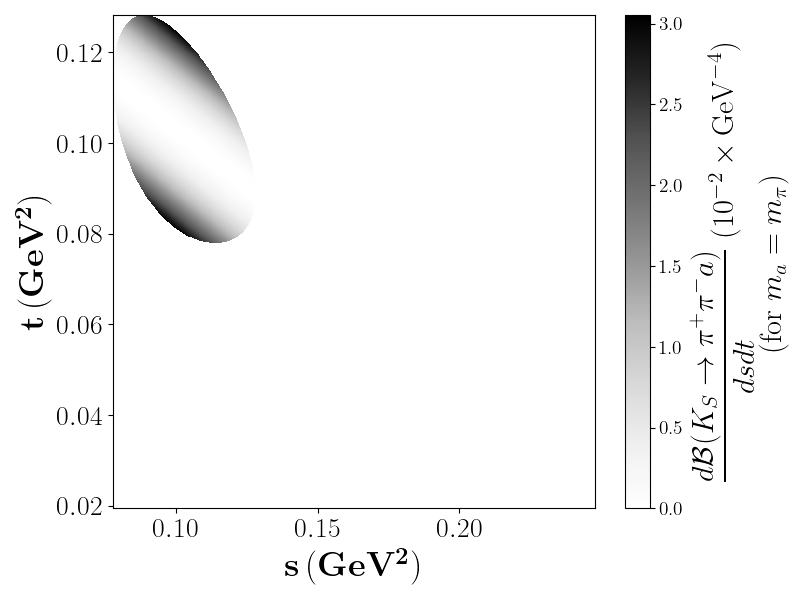}
\centering
\caption{Same as \reffig{fig:dBR_dsdt_Kp}, but for the contribution to $\BR(K_S \to \pi^+\pi^- a)$ proportional to $\re(k_A)_{23}$.
	}
\label{fig:dBR_dsdt_KS_pipi_real}
\end{figure}

\begin{figure}[h!]
\centering
\includegraphics[width=0.49\textwidth]{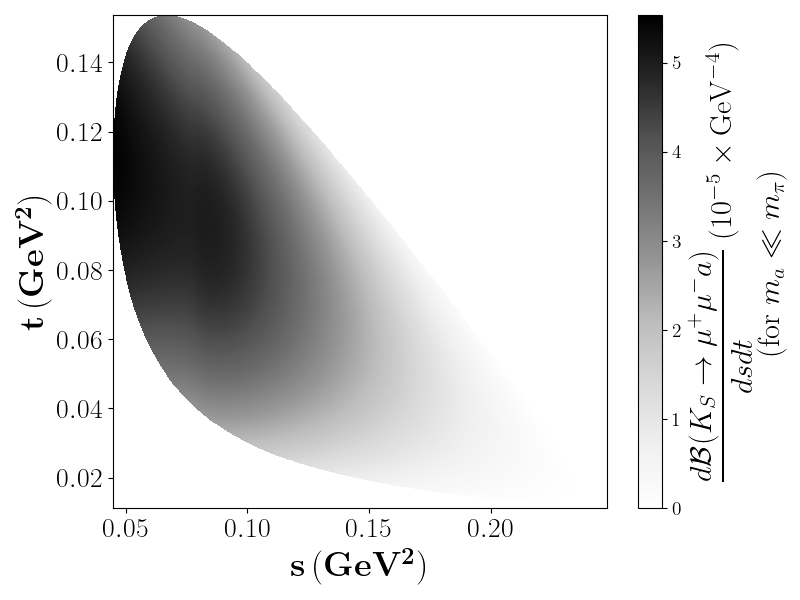}
\includegraphics[width=0.49\textwidth]{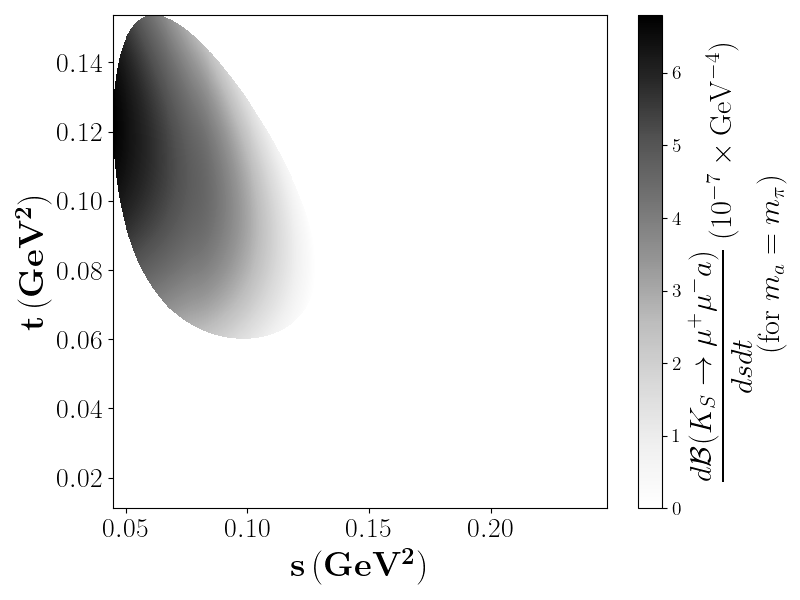}
\centering
\caption{Same as \reffig{fig:dBR_dsdt_Kp} but for $\BR(K_S \to \mu^+\mu^- a)$.}
\label{fig:dBR_dsdt_KS_mumu}
\end{figure}

\begin{figure}[h!]
\centering
\includegraphics[width=0.75\textwidth]{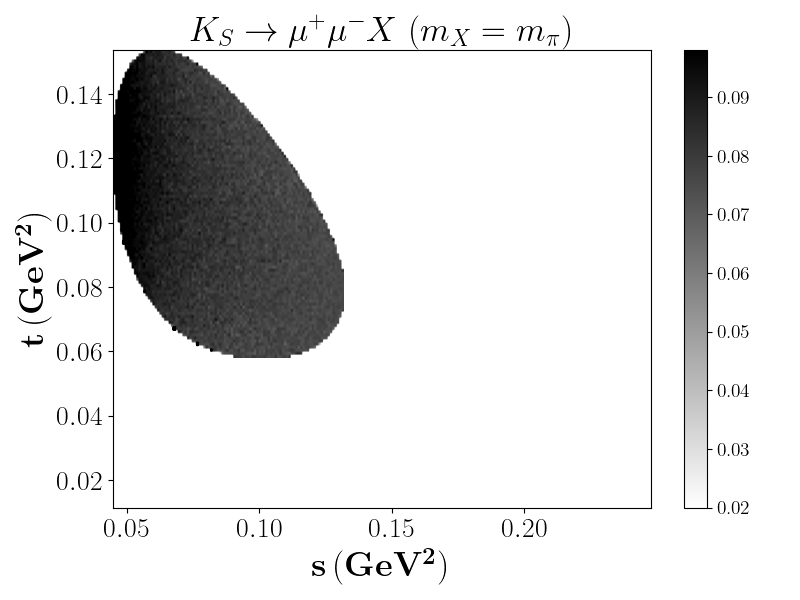}
\centering
\caption{Example of a reconstruction efficiency map, obtained through the fast-simulation code \cite{\fastsim} as detailed in the text. This example corresponds to the case of $K_S \to \mu^+ \mu^- X$ with $m_X = m_\pi$, where $X$ is a generic spinless particle, i.e. not necessarily an axion.}
\label{fig:efficiency_KS_mumu}
\end{figure}

In short, especially for the processes in \reffigs{fig:dBR_dsdt_Kp}{fig:dBR_dsdt_KS_pipi_real}, the $s$ vs. $t$ plane is primarily populated in very specific regions. An interesting question is whether the highest reconstruction efficiencies, as a function of $s, t$, can realistically be found in these most populated regions. Addressing this question in full in a given channel amounts to performing a full-fledged search in that channel. Here we limit ourselves to exploring this question with the fast-simulation tool described in Ref.~\cite{\fastsim}, and designed to estimate the performance of an LHCb-like detector. We thus focus on the $K_S \to \pippim a$ channel in \reffigs{fig:dBR_dsdt_KS_pipi_real}{fig:dBR_dsdt_KS_pipi_imag}, and on the $K_S \to \mupmum a$ channel in \reffig{fig:dBR_dsdt_KS_mumu}, namely on the two channels that can be considered at LHCb. We produce proton-proton collisions according to the present LHC setup, and sample reconstructed events  of the kind $K_S \to \{\pippim, \mupmum\} $ plus a neutral spinless particle. As described in detail in Ref.~\cite{\fastsim}, the detector simulation is an improved version of the fast simulation framework used in Ref.~\cite{\alves}. We then plot, in $(s,t)$-plane bins, the number of events where the two charged final states are correctly reconstructed, normalized to the total yield in the same bin. In \reffig{fig:efficiency_KS_mumu} we provide an explicit example for the di-muon channel and for $m_a = m_\pi$. This plot shows a pattern of efficiencies monotonically decreasing with increasing $s$ and roughly insensitive to $t$. Comparing with the rightmost panel of \reffig{fig:dBR_dsdt_KS_mumu}, this plot suggests that regions with larger efficiencies can be naturally made to match regions with larger yields. Similar considerations can be made for all the other cases.

\section{Outlook} \label{sec:outlook}

We examine three-body kaon decays into $\pi \pi a$ or $\mu \mu a$. First, we perform a complete theoretical study within the context of ChPT plus an axion. We then use this analysis to carry out a sensitivity study of the ratio between the fundamental axial axion-down-strange coupling and the axion decay constant $f_a$. We estimate the sensitivity channel by channel by identifying the respective irreducible backgrounds, and using their associated uncertainties. Using these sensitivity estimates, we find that targeted searches could probe $f_a$ scales in the range of $10^4$–$10^6$ TeV, underscoring their high potential. We validate this conclusion in the case of $K^+ \to \pi^+ \pi^0 a$, for which a recent search is available. In this case, we are able to produce an actual limit, see \refeq{eq:bound_Kp}. This comparison also suggests that our approach yields conservative sensitivities, implying that dedicated searches in the channels considered may perform significantly better than projected. If this is the case, three-body $K$ decays may become the golden channels for constraining $(k_A)_{23}$.

\section*{Acknowledgments}
We thank Alexandre Carvunis and Camille Normand for involvement in the initial stages of this project. We are grateful to Cristina Lazzeroni and Joel Swallow for useful input. We would like to acknowledge the Mainz Institute for Theoretical Physics (MITP) of the Cluster of Excellence PRISMA+ (Project ID 390831469) for enabling us to complete the final stages of this work. This research has received funding from the French ANR, under contracts ANR-19-CE31-0016 (`GammaRare') and ANR-23-CE31-0018 (`InvISYble'), that we gratefully acknowledge.

\bibliographystyle{JHEP}
\bibliography{bibliography}

\end{document}